\documentclass[twocolumn]{aastex61}
\usepackage{amsmath}
\usepackage{verbatim}
\accepted{2019 February 18}
\submitjournal{The Astrophysical Journal}

\shorttitle{S\'ersic Index v/s Effective Radius relation} 
\shortauthors{Marchi-Lasch et al.}

\newcommand{\muMv}{\ifmmode \mu_\mathrm{0}-M_\mathrm{V}\else $\mu_\mathrm{0}-M_\mathrm{V}$\fi}
\newcommand{\nsMv}{\ifmmode n-M_\mathrm{V}\else $n-M_\mathrm{V}$\fi}
\newcommand{\nsRe}{\ifmmode n - R_\mathrm{e}\else $n-R_\mathrm{e}$\fi}
\newcommand{\centmu}{\ifmmode \mu_\mathrm{0}\else $\mu_\mathrm{0}$\fi}
\newcommand{\effmu}{\ifmmode \mu_\mathrm{e}\else $\mu_\mathrm{e}$\fi}
\newcommand{\centVmu}{\ifmmode \mu_\mathrm{V,0}\else $\mu_\mathrm{V,0}$\fi}
\newcommand{\effVmu}{\ifmmode \mu_\mathrm{V,e}\else $\mu_\mathrm{V,e}$\fi}
\newcommand{\Mv}{\ifmmode M_\mathrm{V}\else $M_\mathrm{V}$\fi}
\newcommand{\effrad}{\ifmmode R_\mathrm{e}\else $R_\mathrm{e}$\fi}
\newcommand{\ns}{\ifmmode n\else $n$\fi}
\newcommand{\ellip}{\ifmmode \epsilon\else $\epsilon$\fi}

\begin{document}
\title{A MegaCam Survey of Outer Halo Satellites. VII. A Single S\'ersic Index v/s Effective Radius Relation for Milky Way Outer Halo Satellites.\textsuperscript{1,2}}

\correspondingauthor{Sebasti\'an Marchi}
\email{smarchi@das.uchile.cl}

\affil{Based on observations obtained at the Canada-France-Hawaii Telescope (CFHT) which is operated by the National Research Council of Canada, the Institut National des Sciences de l'Univers of the Centre National de la Recherche Scientifique of France, and the University of Hawaii.}

\affil{This paper includes data gathered with the 6.5 meter Magellan Telescopes located at Las Campanas Observatory, Chile.}

\author{Sebasti\'an Marchi-Lasch}
\affil{Departamento de Astronom\'ia, Universidad de Chile, Camino el Observatorio 1515, Las Condes, Santiago, Chile}

\author{Ricardo R. Mu\~noz}
\affil{Departamento de Astronom\'ia, Universidad de Chile, Camino el Observatorio 1515, Las Condes, Santiago, Chile}

\author{Felipe A. Santana}
\affil{Departamento de Astronom\'ia, Universidad de Chile, Camino el Observatorio 1515, Las Condes, Santiago, Chile}

\author{Julio A. Carballo-Bello}
\affil{Instituto de Astrof\'isica, Facultad de F\'isica, Pontificia Universidad Cat\'olica de Chile, Av. Vicu\~na Mackenna 4860, 782-0436 Macul, Santiago, Chile}

\author{Julio Chanam\'e}
\affil{Instituto de Astrof\'isica, Facultad de F\'isica, Pontificia Universidad Cat\'olica de Chile, Av. Vicu\~na Mackenna 4860, 782-0436 Macul, Santiago, Chile}

\author{Marla Geha}
\affil{Astronomy Department, Yale University, New Haven, CT 06520, USA}

\author{Joshua D. Simon}
\affil{Observatories of the Carnegie Institution for Science, 813 Santa Barbara St., Pasadena, CA 91101, USA}

\author{Peter B. Stetson}
\affil{National Research Council of Canada, Herzberg Astronomy and Astrophysics, 5071 W. Saanich Road, Victoria, BC V9E 2E7, Canada}

\author{S. G. Djorgovski}
\affil{Astronomy Department, California Institute of Technology, Pasadena, CA, 91125, USA}

\begin{abstract}
In this work we use structural properties of Milky Way's outer
halo ($R_G > 25\,\mathrm{kpc}$) satellites (dwarf spheroidal galaxies, ultra-faint dwarf galaxies and globular clusters) derived from deep, wide-field and homogeneous data, to present evidence of a correlation in the S\'ersic index v/s effective radius plane followed by a large fraction of outer halo globular clusters and satellite dwarf galaxies. We show that this correlation can be entirely reproduced by fitting empirical relations in the central surface brightness v/s absolute magnitude and S\'ersic index v/s absolute magnitude parameter spaces, and by assuming the existence of two types of outer halo globular clusters: one of high surface brightness (HSB group), with properties similar to inner halo clusters; and another of low surface brightness (LSB group), which share characteristics with dwarf spheroidal and ultra-faint dwarf galaxies. 
Given the similarities of LSB clusters with dwarf spheroidal and ultra-faint dwarf galaxies, we discuss the possibility that outer halo clusters also originated inside dark matter halos and that tidal forces from different galaxy host's potentials are responsible for the different properties between HSB and LSB clusters.
\end{abstract}


\section{Introduction} \label{introduction}
The halo of the Milky Way (MW) contains important information about the ancient history of our Galaxy, especially since dynamical scales are long enough to retain information of past Galactic events  \citep[e.g.,][]{Johnston1996,Mayer2002}. 
A significant fraction of this information is contained in the structural, dynamical and chemical properties of satellite stellar structures of the MW, which dominate the outer halo stellar distribution \citep[see][for reviews on this topic]{Majewski2004,Willman2010,Ivezic2012}. 
Thus, by studying these substructures it is, in principle, possible to peer into our galaxy's past and learn about the processes that governed its formation and evolution. 

The stellar structures that surround the Galaxy have been usually classified as either globular clusters (GCs) or 
dwarf galaxies. Specifically, most of the dwarf galaxies are of the dwarf spheroidal type (dSph), which are devoid of gas and show no current stellar formation. Both types of stellar structures are dominated by an old, metal-poor stellar population. Currently, it is accepted that dwarf galaxies formed their own Dark Matter (DM) halos at small scales and were accreted later by the MW, as described in hierarchical growth models \citep{Searle1978,Bullock2005}. In the case of GCs, a fraction of them formed together with our galaxy during a phase of rapid collapse as proposed by \citet{Eggen1962}, whereas others are thought to have an external origin, i.e. they formed in galaxies that were later accreted by the MW, which stripped off their GCs \citep{Zinn1993,Zinn1996,Mackey2004,Mackey2005,Leaman2013,Zaritsky2016}.

To understand better the role of these structures in the formation and evolution of the MW, current research efforts have focused on the detection of satellites in order to obtain a reliable census of satellite objects orbiting our Galaxy. The results have significantly changed the way we understand our Galaxy surroundings. Before 2005, only nine Galactic dSphs were known (now referred to as classical dSphs), with luminosities in the range  $-12 \lesssim M_\mathrm{V} \lesssim -8$ and with half-light radii on the order of $100\,\mathrm{pc}$. Regarding GCs, almost all of them were compact objects, with half-light radii of less than $10\,\mathrm{pc}$ and, in general, less luminous than classical dSphs. Over the last decade and a half and thanks to large area surveys like the Sloan Digital Sky Survey 
\citep[SDSS,][]{York2000}, PanSTARRS1 \citep{Chambers2016} and the 
Dark Energy Survey \citep[DES,][]{DESCol2016}, the population of satellite systems has increased significantly, more than doubling the total number \citep{Willman2005,Zucker2006,Belokurov2006,Belokurov2007,Belokurov2008,Belokurov2009,Belokurov2010,Belokurov2014,Walsh2007,Munoz2012a,Bechtol2015,Kim2015a,Koposov2015,Laevens2015a,Laevens2015b,Martin2015,Torrealba2016a,Torrealba2016b,Homma2017}. The new objects include low-luminosity dSphs ($M_\mathrm{V} > -8$), named ultra-faint dwarf galaxies (UFDs), some of them as small as some GCs; and halo GCs, some of them of size comparable to these UFDs. In this new scenario, the size gap that seemed to separate GCs from dSphs in the size v/s luminosity plot has started to populate, casting doubts about the true different origins for extended GCs and UFDs \citep[e.g.,][]{Drlica-Wagner2015,Torrealba2016b}. For example, it is not clear whether the large half-light radii of extended GCs is an intrinsic property of a different class of object or a result of interactions with the MW \citep[e.g.,][]{vandenBergh2004, Ripepi2007, Hwang2011}.

One widely accepted difference between dSphs and GCs is their DM content. For classical dSphs, the mass-to-light ratio within their half-light radii ranges from $\sim 6$ to $\sim 100$, and from $\sim 100$ to $\sim 3000$ for UFDs (see Figure 11 of \cite{mcconnachie2012}), making the latter the most DM dominated objects known in the Universe. In the case of GCs, they have values consistent with no DM content, with typical values for the mass-to-light ratio of $\sim$ 1 to $\sim$ 4 \citep[e.g.,][]{McLaughlin2000, Rejkuba2007, Baumgardt2009}. This feature, together with the different metallicity spread between dSphs and GCs, has become the standard to classify a halo stellar overdensity either as a dSph or a GC \citep{Willman2012}. In principle, it is reasonable to think that the presence of DM should leave a distinct imprint in the structural and photometric parameters of the baryonic matter of dSphs, in stark contrast with GCs. To explore this idea and to shed some light into the different processes that formed this two type of substructures, it is useful to have a complete characterization of their respective structural and photometric properties and to compare them homogeneously. 

This paper is part of a series of articles based on a
catalog of structural parameters constructed from deep, wide and homogeneous observations of 58 satellite objects located in the outer halo of the MW
\citep[][]{munoz17a, munoz17b}. These parameters include half-light radius, surface brightness, luminosity, ellipticity and S\'ersic index.
In C\^ot\'e et al. (in preparation) we study a wide range of scaling relations between the different objects in the
catalog. In this article we focus on the observed trend of S\'ersic index with effective radius, which shows a strong correlation when all outer halo objects are considered.

This article is organized as follows. In Section \ref{data}, we briefly describe our dataset. In Section \ref{distros}, we use the structural parameters derived from our dataset to explore some properties of GCs and dSphs, while in Section \ref{correlation} we concentrate our analysis in the Sersic's index v/s effective radius relation. Next, in Section \ref{discussion} we provide an explanation for the origin of the previous correlation and discuss some consequences of its existence for formation and evolution processes of GCs and dSphs. Finally, in Section \ref{conclusions} we present a summary and the conclusions of this work.

\section{Data} \label{data}

The dataset used in this work is composed by observations of 58 satellite objects of the MW, which includes GCs, classical dSphs and UFDs and a number of objects not yet classified (i.e., their structural properties do not allow for a clear differentiation nor their DM content or metallicity spread are known). The classification for each object is based on information in the literature.

Observations of 44 of these objects were carried out using the MegaCam imager on the Canada-France-Hawaii Telescope (CFHT) in the northern hemisphere and the Megacam imager on the Magellan II-Clay Telescope at Las Campanas Observatory in the southern hemisphere. The  data for the remaining 14 objects were obtain from different sources, most of them from public data from the DES Year Release 1 \cite[see][for details]{munoz17a}.

\citet{munoz17a} describes the data reduction, astrometry, point source photometry and photometric calibration  performed to the whole sample, in order to obtain a homogeneous dataset. To measure the structural parameters and density profiles, a maximum likelihood approach was applied to the observations of every object, assuming a S\'ersic density profile \citep{Sersic1968} plus a background density \citep{munoz17b}. Absolute magnitude and surface brightness values were obtained by integrating a theoretical luminosity function for every object, which is normalized by the object's number of member stars.

It is important to mention that, although our observations come from different instruments and full photometric homogeneity is not possible, as described in \citet{munoz17a} care was taken to make the dataset as homogeneous as possible: the Megacam imagers used to create the primary catalog are similar in structure and performance, the same bands were used for all 58 objects, the same reduction pipeline and techniques were used for all objects, the spatial coverage for every object in our dataset is comparable  (at least 5 effective radii) with only a few exceptions.

The outer halo object data used in this work is presented in Table \ref{tab:outer_objs}.

\startlongtable
\begin{deluxetable*}{llrrrrrr}
\tablecaption{Outer halo object parameters used in this work. \label{tab:outer_objs}}
\tablewidth{0pt}
\tablehead{
\colhead{Object} &
\colhead{Type} &
\colhead{\Mv} & 
\colhead{\centVmu} &
\colhead{\effVmu} &
\colhead{\effrad} & 
\colhead{\ns} &
\colhead{\ellip}\\
\colhead{} & 
\colhead{} &
\colhead{} & 
\colhead{(mag$/\arcsec^2$)} & 
\colhead{(mag$/\arcsec^2$)} &
\colhead{(pc)} & 
\colhead{} &
\colhead{}\\
}
\decimals
\startdata
AM 1 & Outer Halo GC & $-5.02 \pm 0.26$ & $23.19^{+0.39}_{-0.40}$ & $25.18^{+0.39}_{-0.40}$ & $16.50 \pm 1.08$ & $1.08 \pm 0.13$ & $0.16 \pm 0.06$ \\
AM 4 & Outer Halo GC & $-0.89 \pm 0.81$ & $24.74^{+1.18}_{-1.25}$ & $27.51^{+1.18}_{-1.25}$ & $7.34 \pm 1.35$ & $1.44 \pm 0.33$ & $0.29 \pm 0.14$ \\
Balbinot 1 & Outer Halo GC & $-1.21 \pm 0.89$ & $24.38^{+1.16}_{-1.20}$ & $27.24^{+1.16}_{-1.20}$ & $7.79 \pm 1.02$ & $1.48 \pm 0.23$ & $0.35 \pm 0.10$ \\
Bootes I & UFD & $-6.00 \pm 0.25$ & $28.40 \pm 0.31$ & $29.43 \pm 0.31$ & $216.18 \pm 5.18$ & $0.64 \pm 0.03$ & $0.25 \pm 0.02$ \\
Bootes II & UFD & $-2.92 \pm 0.74$ & $27.56^{+1.04}_{-1.08}$ & $28.75^{+1.04}_{-1.08}$ & $37.26 \pm 5.50$ & $0.71 \pm 0.43$ & $0.24 \pm 0.12$ \\
CVn I & UFD & $-8.48 \pm 0.13$ & $27.10 \pm 0.19$ & $28.44 \pm 0.19$ & $486.38 \pm 14.59$ & $0.78 \pm 0.04$ & $0.46 \pm 0.02$ \\
CVn II & UFD & $-4.85 \pm 0.36$ & $26.83^{+0.67}_{-0.72}$ & $27.76^{+0.67}_{-0.72}$ & $70.28 \pm 10.70$ & $0.59 \pm 0.49$ & $0.46 \pm 0.11$ \\
Carina & dSph & $-9.42 \pm 0.05$ & $25.27 \pm 0.07$ & $26.74 \pm 0.07$ & $312.76 \pm 3.36$ & $0.84 \pm 0.02$ & $0.37 \pm 0.01$ \\
ComBer & UFD & $-4.36 \pm 0.25$ & $26.99^{+0.36}_{-0.37}$ & $28.66^{+0.36}_{-0.37}$ & $72.06 \pm 3.84$ & $0.93 \pm 0.12$ & $0.37 \pm 0.05$ \\
Draco & dSph & $-8.70 \pm 0.05$ & $25.01 \pm 0.07$ & $26.74 \pm 0.07$ & $207.15 \pm 1.99$ & $0.96 \pm 0.02$ & $0.30 \pm 0.01$ \\
Eridanus & Outer Halo GC & $-4.92 \pm 0.26$ & $23.24 \pm 0.40$ & $25.45 \pm 0.40$ & $16.77 \pm 1.05$ & $1.18 \pm 0.14$ & $0.09 \pm 0.04$ \\
Eridanus II & UFD & $-7.19 \pm 0.09$ & $26.64^{+0.29}_{-0.31}$ & $27.96^{+0.29}_{-0.31}$ & $200.07 \pm 18.79$ & $0.77 \pm 0.19$ & $0.37 \pm 0.06$ \\
Eridanus III & Not classified & $-7.19 \pm 0.09$ & $18.01^{+1.36}_{-3.51}$ & $21.22^{+1.36}_{-3.51}$ & $7.34 \pm 5.82$ & $1.64 \pm 0.27$ & $0.32 \pm 0.13$ \\
Fornax & dSph & $-13.45 \pm 0.14$ & $23.60^{+0.16}_{-0.17}$ & $24.79^{+0.16}_{-0.17}$ & $786.80 \pm 8.55$ & $0.71 \pm 0.01$ & $0.28 \pm 0.01$ \\
Grus 1 & UFD & $-3.46 \pm 0.59$ & $26.87^{+1.35}_{-1.76}$ & $29.41^{+1.35}_{-1.76}$ & $72.61 \pm 30.37$ & $1.33 \pm 0.31$ & $0.54 \pm 0.26$ \\
Hercules & UFD & $-5.19 \pm 0.45$ & $27.47^{+0.65}_{-0.67}$ & $29.70^{+0.65}_{-0.67}$ & $230.00 \pm 22.27$ & $1.19 \pm 0.17$ & $0.69 \pm 0.04$ \\
Horologium I & UFD & $-3.53 \pm 0.56$ & $26.29^{+0.99}_{-1.10}$ & $28.07^{+0.99}_{-1.10}$ & $35.39 \pm 7.81$ & $0.98 \pm 0.47$ & $0.31 \pm 0.16$ \\
Horologium II & UFD & $-1.54 \pm 1.02$ & $27.66^{+1.85}_{-2.37}$ & $29.67^{+1.85}_{-2.37}$ & $64.21 \pm 29.72$ & $1.09 \pm 0.37$ & $0.86 \pm 0.19$ \\
Hydra II & UFD & $-4.58 \pm 0.37$ & $26.15^{+0.79}_{-0.89}$ & $28.40^{+0.79}_{-0.89}$ & $58.47 \pm 12.47$ & $1.20 \pm 0.46$ & $0.17 \pm 0.13$ \\
Indus 1 & Not classified & $-3.31 \pm 0.62$ & $24.39^{+1.53}_{-2.20}$ & $26.69^{+1.53}_{-2.20}$ & $25.31 \pm 13.09$ & $1.22 \pm 0.44$ & $0.72 \pm 0.29$ \\
Koposov 1 & Outer Halo GC & $-1.03 \pm 0.69$ & $25.11^{+1.18}_{-1.32}$ & $27.51^{+1.18}_{-1.32}$ & $10.12 \pm 2.53$ & $1.27 \pm 0.56$ & $0.55 \pm 0.15$ \\
Koposov 2 & Outer Halo GC & $-0.91 \pm 0.81$ & $23.40^{+1.22}_{-1.32}$ & $25.98^{+1.22}_{-1.32}$ & $4.34 \pm 0.91$ & $1.35 \pm 0.70$ & $0.48 \pm 0.12$ \\
Laevens 1 & Outer Halo GC & $-4.79 \pm 0.33$ & $24.50^{+0.62}_{-0.67}$ & $25.81^{+0.62}_{-0.67}$ & $20.67 \pm 2.95$ & $0.77 \pm 0.36$ & $0.11 \pm 0.10$ \\
Laevens 2 & UFD & $-1.59 \pm 0.76$ & $25.74^{+1.15}_{-1.24}$ & $28.53^{+1.15}_{-1.24}$ & $17.45 \pm 3.49$ & $1.45 \pm 0.45$ & $0.39 \pm 0.11$ \\
Leo I & dSph & $-11.76 \pm 0.28$ & $22.62 \pm 0.30$ & $23.93 \pm 0.30$ & $243.82 \pm 2.22$ & $0.77 \pm 0.02$ & $0.30 \pm 0.01$ \\
Leo II & dSph & $-9.73 \pm 0.04$ & $24.25 \pm 0.06$ & $25.43 \pm 0.06$ & $168.09 \pm 2.03$ & $0.71 \pm 0.02$ & $0.07 \pm 0.02$ \\
Leo IV & UFD & $-4.98 \pm 0.26$ & $27.82^{+0.51}_{-0.54}$ & $29.33^{+0.51}_{-0.54}$ & $116.92 \pm 13.89$ & $0.86 \pm 0.26$ & $0.19 \pm 0.09$ \\
Leo T & UFD & $-7.59 \pm 0.14$ & $25.43^{+0.37}_{-0.40}$ & $27.31^{+0.37}_{-0.40}$ & $151.63 \pm 16.98$ & $1.03 \pm 0.26$ & $0.23 \pm 0.09$ \\
Leo V & UFD & $-4.39 \pm 0.36$ & $24.90^{+0.79}_{-0.90}$ & $28.24^{+0.79}_{-0.90}$ & $51.78 \pm 11.39$ & $1.70 \pm 0.36$ & $0.35 \pm 0.07$ \\
Mu\~noz 1 & Outer Halo GC & $-0.48 \pm 0.97$ & $26.34^{+1.34}_{-1.42}$ & $30.08^{+1.34}_{-1.42}$ & $22.25 \pm 4.19$ & $1.89 \pm 0.31$ & $0.50 \pm 0.05$ \\
NGC 2419 & Outer Halo GC & $-9.33 \pm 0.03$ & $18.83 \pm 0.05$ & $22.19 \pm 0.05$ & $25.71 \pm 0.24$ & $1.71 \pm 0.02$ & $0.05 \pm 0.01$ \\
NGC 5694 & Outer Halo GC & $-7.93 \pm 0.09$ & $13.42 \pm 0.14$ & $20.01 \pm 0.14$ & $4.28 \pm 0.10$ & $3.20 \pm 0.08$ & $0.06 \pm 0.02$ \\
NGC 5824 & Outer Halo GC & $-9.28 \pm 0.04$ & $11.15 \pm 0.08$ & $19.09 \pm 0.08$ & $4.95 \pm 0.09$ & $3.82 \pm 0.05$ & $0.04 \pm 0.01$ \\
NGC 6229 & Outer Halo GC & $-8.03 \pm 0.16$ & $13.88 \pm 0.22$ & $19.21 \pm 0.22$ & $3.19 \pm 0.09$ & $2.62 \pm 0.08$ & $0.02 \pm 0.01$ \\
NGC 7006 & Outer Halo GC & $-7.41 \pm 0.08$ & $15.99 \pm 0.13$ & $21.17 \pm 0.13$ & $6.11 \pm 0.12$ & $2.55 \pm 0.07$ & $0.07 \pm 0.01$ \\
NGC 7492 & Outer Halo GC & $-6.10 \pm 0.04$ & $21.24 \pm 0.06$ & $23.05 \pm 0.06$ & $9.56 \pm 0.08$ & $1.00 \pm 0.02$ & $0.02 \pm 0.02$ \\
Palomar 13 & Outer Halo GC & $-2.82 \pm 0.55$ & $22.15^{+0.70}_{-0.71}$ & $26.61^{+0.70}_{-0.71}$ & $9.53 \pm 0.68$ & $2.22 \pm 0.19$ & $0.10 \pm 0.06$ \\
Palomar 14 & Outer Halo GC & $-5.39 \pm 0.24$ & $23.59 \pm 0.33$ & $26.47 \pm 0.33$ & $32.04 \pm 1.34$ & $1.49 \pm 0.08$ & $0.11 \pm 0.04$ \\
Palomar 15 & Outer Halo GC & $-5.65 \pm 0.19$ & $23.07 \pm 0.24$ & $24.97 \pm 0.24$ & $19.02 \pm 0.39$ & $1.04 \pm 0.06$ & $0.05 \pm 0.02$ \\
Palomar 2 & Outer Halo GC & $-9.05 \pm 0.07$ & $16.57^{+0.11}_{-0.12}$ & $19.88^{+0.11}_{-0.12}$ & $7.83 \pm 0.16$ & $1.69 \pm 0.04$ & $0.05 \pm 0.02$ \\
Palomar 3 & Outer Halo GC & $-5.48 \pm 0.21$ & $23.55^{+0.27}_{-0.28}$ & $25.08^{+0.27}_{-0.28}$ & $19.37 \pm 0.54$ & $0.87 \pm 0.05$ & $0.07 \pm 0.03$ \\
Palomar 4 & Outer Halo GC & $-6.01 \pm 0.16$ & $22.74^{+0.22}_{-0.23}$ & $24.81^{+0.22}_{-0.23}$ & $20.24 \pm 0.63$ & $1.12 \pm 0.08$ & $0.03 \pm 0.02$ \\
Phoenix 2 & Not classified & $-3.28 \pm 0.63$ & $25.85^{+0.97}_{-1.03}$ & $27.97^{+0.97}_{-1.03}$ & $38.87 \pm 6.52$ & $1.14 \pm 0.27$ & $0.61 \pm 0.15$ \\
Pictoris 1 & Not classified & $-3.44 \pm 0.60$ & $24.51^{+1.46}_{-2.04}$ & $27.43^{+1.46}_{-2.04}$ & $21.89 \pm 10.61$ & $1.51 \pm 0.31$ & $0.24 \pm 0.19$ \\
Pisces II & UFD & $-4.21 \pm 0.38$ & $26.53^{+0.71}_{-0.77}$ & $28.61^{+0.71}_{-0.77}$ & $64.59 \pm 10.59$ & $1.12 \pm 0.34$ & $0.40 \pm 0.10$ \\
Pyxis & Outer Halo GC & $-5.69 \pm 0.19$ & $23.07^{+0.24}_{-0.25}$ & $24.87^{+0.24}_{-0.25}$ & $18.57 \pm 0.46$ & $0.99 \pm 0.05$ & $0.04 \pm 0.02$ \\
Reticulum II & UFD & $-3.86 \pm 0.38$ & $26.79 \pm 0.46$ & $27.73 \pm 0.46$ & $48.78 \pm 1.83$ & $0.60 \pm 0.05$ & $0.56 \pm 0.03$ \\
Sculptor & dSph & $-10.81 \pm 0.14$ & $23.41^{+0.36}_{-0.38}$ & $24.66^{+0.36}_{-0.38}$ & $215.14 \pm 22.51$ & $0.74 \pm 0.07$ & $0.26 \pm 0.01$ \\
Segue 1 & UFD & $-1.29 \pm 0.73$ & $28.08^{+0.98}_{-1.01}$ & $29.57^{+0.98}_{-1.01}$ & $26.43 \pm 3.21$ & $0.85 \pm 0.28$ & $0.34 \pm 0.11$ \\
Segue 2 & UFD & $-1.85 \pm 0.88$ & $28.49^{+1.05}_{-1.06}$ & $29.92^{+1.05}_{-1.06}$ & $37.06 \pm 2.95$ & $0.82 \pm 0.16$ & $0.21 \pm 0.07$ \\
Segue 3 & Outer Halo GC & $-0.85 \pm 0.67$ & $23.86^{+1.02}_{-1.08}$ & $26.32^{+1.02}_{-1.08}$ & $4.08 \pm 0.71$ & $1.30 \pm 0.30$ & $0.22 \pm 0.09$ \\
Sextans & dSph & $-8.71 \pm 0.06$ & $27.23 \pm 0.08$ & $28.18 \pm 0.08$ & $442.04 \pm 4.25$ & $0.60 \pm 0.01$ & $0.30 \pm 0.01$ \\
UMa I & UFD & $-5.12 \pm 0.38$ & $29.12^{+0.47}_{-0.48}$ & $29.78^{+0.47}_{-0.48}$ & $235.32 \pm 9.59$ & $0.47 \pm 0.08$ & $0.57 \pm 0.03$ \\
UMa II & UFD & $-4.23 \pm 0.26$ & $28.08 \pm 0.33$ & $29.66 \pm 0.33$ & $129.85 \pm 4.28$ & $0.89 \pm 0.10$ & $0.56 \pm 0.03$ \\
UMi & dSph & $-9.02 \pm 0.05$ & $25.77 \pm 0.06$ & $27.09 \pm 0.06$ & $367.21 \pm 2.43$ & $0.77 \pm 0.01$ & $0.55 \pm 0.01$ \\
Whiting 1 & Outer Halo GC & $-2.54 \pm 0.44$ & $21.45^{+0.64}_{-0.66}$ & $25.84^{+0.64}_{-0.66}$ & $6.39 \pm 0.61$ & $2.19 \pm 0.26$ & $0.24 \pm 0.05$ \\
Willman I & UFD & $-2.52 \pm 0.74$ & $25.88^{+0.92}_{-0.94}$ & $28.43^{+0.92}_{-0.94}$ & $27.97 \pm 2.43$ & $1.34 \pm 0.20$ & $0.47 \pm 0.06$ \\
\enddata
\end{deluxetable*}

\section{Parameters distributions} \label{distros}

\begin{figure*}[ht!]
\plotone{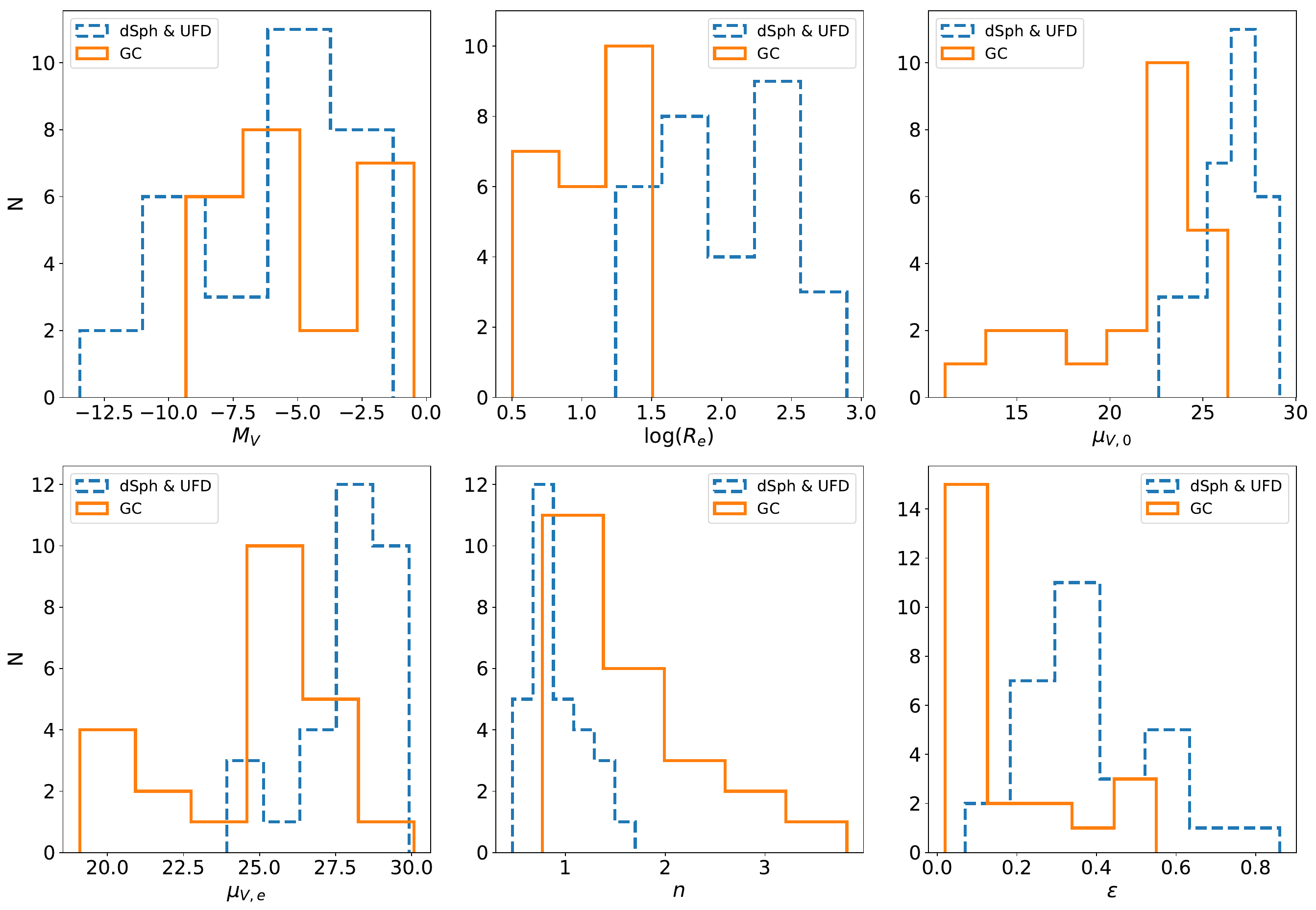}
\caption{Comparison of GCs's and dwarf dSphs's parameter distributions, for all the structural and photometric parameters analyzed in our study. \label{fig:1D_distros}}
\end{figure*}

Giving the characteristics of our new dataset (wider, deeper and nearly homogeneous), in C\^ot\'e et al. (in preparation) we explore in depth a wide range of correlations between different structural parameters in order to globally assess similarities and differences between GCs and dSphs. Here, we briefly highlight some of those results.

Figure \ref{fig:1D_distros} shows the distribution of six structural parameters: absolute magnitude in the V band (\Mv), effective radius (\effrad), central surface brightness in the V band (\centVmu), effective surface brightness in the V band (\effVmu), S\'ersic index (\ns) and ellipticity (\ellip), divided into dwarf galaxies and GCs subgroups. 
In general, as a group dwarf galaxies are larger, brighter, more diffuse, less concentrated and more elongated than GCs, but all parameters show overlap between the two classes of objects, with no clear boundaries separating the two families.

One interesting result from Figure \ref{fig:1D_distros} is the ellipticity distribution (bottom right panel). The vast majority of GC are significantly round, with their ellipcities concentrated around $\ellip < 0.15$. Dwarf galaxies, on the other hand, are distributed along the whole range, preferentially at $\ellip > 0.2$, with the exception of Leo~II, which shows little elongation. However, some GCs extend the distribution to significantly higher ellipticity values, overlapping with most of the dwarf galaxy distribution. In Figure \ref{fig:ellmag} we show how the ellipticity behaves as a function of luminosity for all objects in our catalog.
Most luminous GCs, up to $\Mv \sim -6$, have ellipticities consistent with little or no elongation. This changes at lower luminosities, where GCs are characterized by progressively increasing ellipticities, up to $\sim 0.7$ for the faintest object.

\subsection{Effect of low number of member stars in measured parameters.}

\begin{figure*}[ht!]
\includegraphics[width = \textwidth]{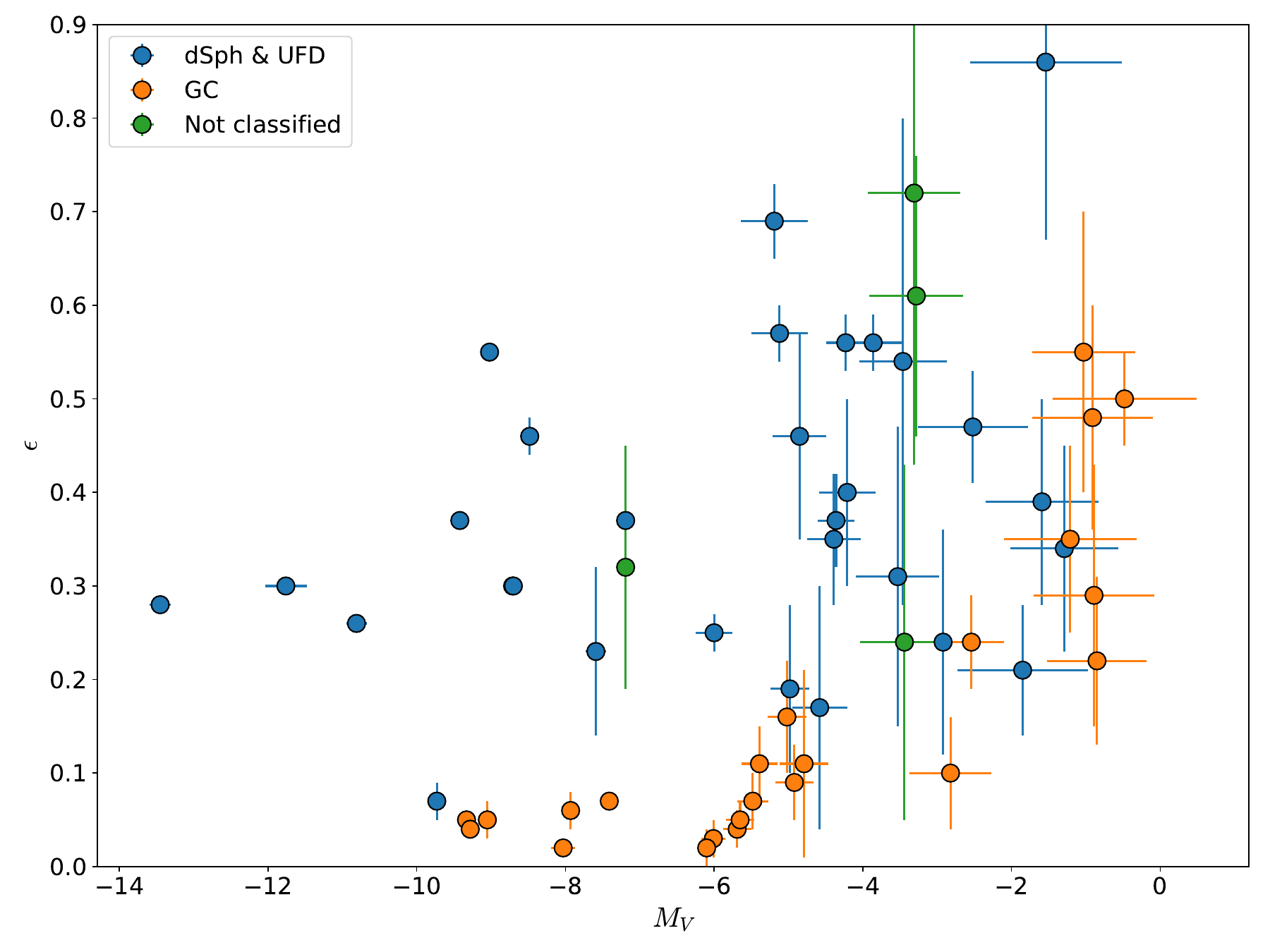}
\caption{Evolution of ellipticity with luminosity for GCs, dSphs and UFDs. GCs with luminosities higher than $\Mv \sim -5$ are clearly more round than dwarf galaxies of comparable luminosity. At lower luminosities, clusters seem to increase their ellipticities to values similar to galaxies's. \label{fig:ellmag}}
\end{figure*}

\begin{figure*}[ht!]
\includegraphics[width = \textwidth]{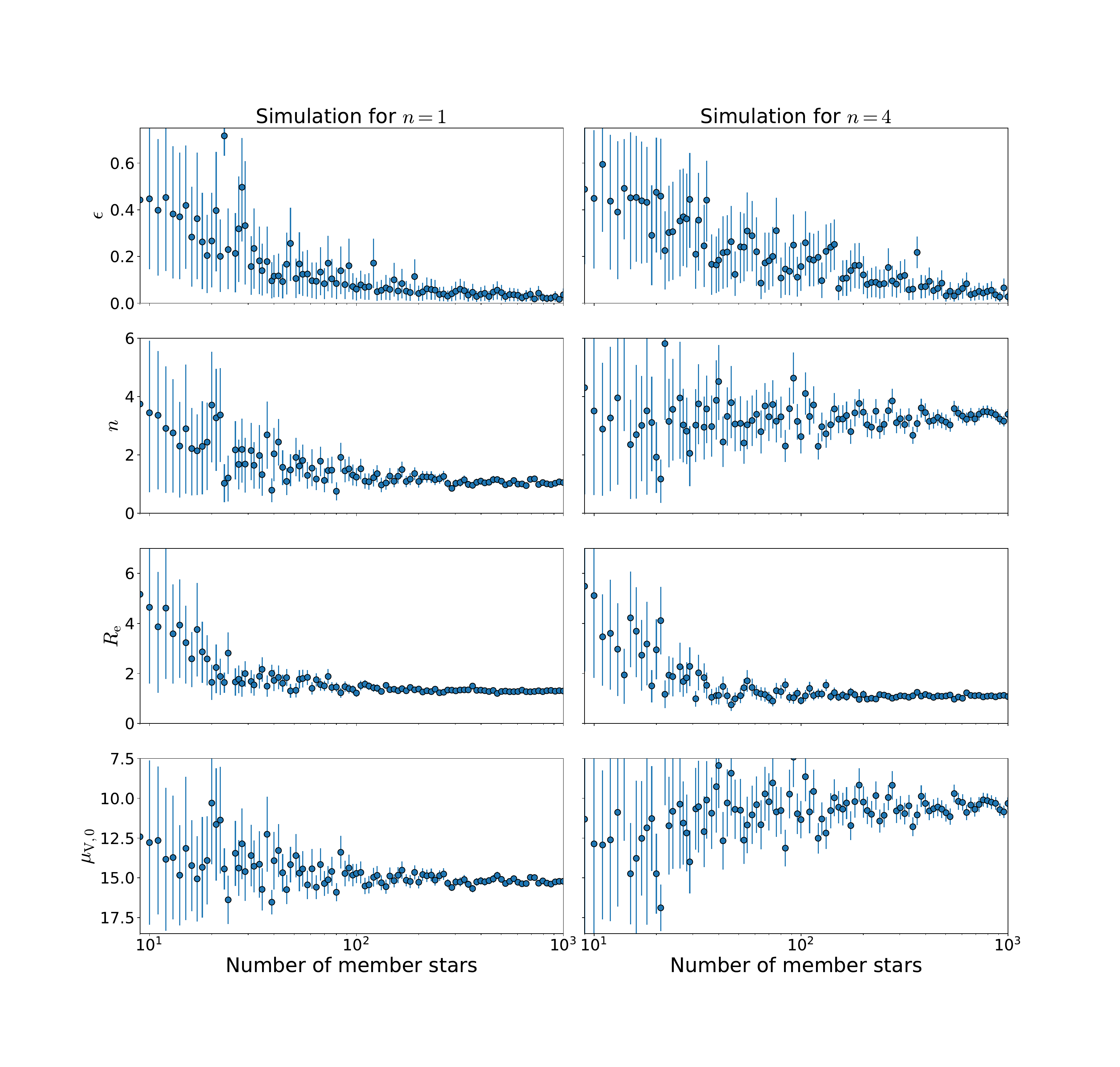}
\caption{Ellipticity, S\'ersic index, effective radius and central surface brightness estimations for different subsamples of member stars from the simulated satellite object. Left panels are the results for $n = 1$, while the right panels are for $n=4$. This plot shows that the increase of ellipticities observed in globular clusters in \ref{fig:ellmag} is likely due to an statistical effect of a low number of observed member stars. This effect is only seen at a much low number of member stars for the rest of the parameters. \label{fig:shotnoise_all_simulated}}
\end{figure*}

\citet{Martin2008b} showed that rounder objects can mistakenly seem elongated if their structural parameters are measured from samples with low numbers of stars.
Similar studies have shown that the low number of stars detected in ultra-low luminosity objects affect our
ability to reliably measure their structural properties
\citep[e.g.,][]{Sand2010, Munoz2012b}.

To explore the effect of a low number of member stars in the measured ellipticities, we fit a stellar density profile over a set of simulated stellar overdensities of different number of member stars. To simulate a stellar overdensity, we generate member stars randomly across a defined field of view using a S\'ersic density profile as a probability distribution. Given that we want to test the potential departure from a round shape at low luminosities, we set the ellipticity of our simulated object equal to 0. We run two sets of simulations, where the effective radius is kept constant at $1.25\arcmin$ and we give two different values to the S\'ersic index, 1 and 4, in order to see the effect of different concentrations. The field of view is given by a $30\, \mathrm{arcmin} \times 30\, \mathrm{arcmin}$ 
area and, for simplicity, we put the simulated object at the center. Finally, we add a particle background given by a constant stellar background density through the whole field of view. The value we adopt for this parameter is $0.5\,[1/\arcmin^2]$, which is a typical value derived by \citet{munoz17b} for the objects in our dataset (see their Figure 4 to 14, right panels).

We generate samples of different number of member stars by taking random subsamples without replacement from the originally simulated object. The number of member star for the subsamples ranges from 1 to 2000 stars in different steps given by a logarithmic scale. For every simulated subsample we then fit a S\'ersic profile with the effective radius, S\'ersic index, central coordinate and ellipticity as free parameters, while the background density is kept fixed. The fit is performed through a bayesian MCMC approach, using the \texttt{emcee} Python package \citep{emcee_ref}. The likelihood function is represented by the S\'ersic density profile and the priors for all the free parameters are defined as uniforms. The density profile used is given by:

\begin{equation}
    \Sigma(r) = \Sigma_{0,\mathrm{S}} \exp \left[-b_n \left(\frac{r}{r_e}\right)^{1/n} \right] + \Sigma_\mathrm{bkg}
\end{equation}

\noindent where $\Sigma(r)$ is the stellar density for any given radius $r$, $\Sigma_{0,\mathrm{S}}$ is the S\'ersic central stellar density, $n$ is the S\'ersic index, $r_e$ is the effective radius, $b_n$ is approximated by $1.999n-0.327$ \citep{Capaccioli1989} and $\Sigma_\mathrm{bkg}$ is the background stellar density. 

The top panels of Figure \ref{fig:shotnoise_all_simulated} show the estimation of the ellipticity for every simulated subsample. It is evident that a low number of member stars increases the bias and the uncertainty of the estimation. This is consistent with the interpretation that the trend in ellipticity that we see in Figure \ref{fig:ellmag} is possibly due to the low luminosity of the satellite objects and it is not a real effect. 

In the same vein, it is likely that the ellipticity measurements of dSphs at low luminosities are also affected by a low number of member stars. Therefore, for absolute magnitudes fainter than $\sim -5$ we cannot reliably use the ellipticity values and thus we cannot clearly establish differences or similarities in ellipticies for GCs and dSphs.

In Figure \ref{fig:shotnoise_all_simulated} we also show the behavior of the S\'ersic index, effective radius and central surface brightness as a function of number of stars. All these parameters appear more robust to shot noise introduced by low number of stars, with significant deviations observed only in the more extreme cases of fewer than $\sim 30$ stars for the $n = 1$ case, and fewer than $\sim 100$ stars for $n = 4$. Although measured parameters for object with true S\'ersic index equal to 4 are more sensitive to the number of member stars, we note that all our objects with high S\'ersic index are dominated be a large number of member stars.  We therefore regard trends involving \effrad, \ns and \centVmu as more reliable, considering that the object with the lowest number of member stars in our sample has on the order of 100 stars.

It is important to mention that the values for \centVmu were not calculated directly from the simulation, but derived as in \citet{munoz17b}. The relations used are:

\begin{equation*}
    I_0 = L b_{\ns}^{2\ns} / [2\pi \effrad^2 \ns \Gamma(2\ns) (1-\ellip)]
\end{equation*}

\begin{equation*}
    \centmu = M_\odot + 21.572 - 2.5 \log I_0
\end{equation*}

\noindent where $b_\ns = 1.9992\ns - 0.3271$, $I_0$ is the central intensity and $L$ is the total luminosity of the object. Uncertainties where calculated by propagation of errors.

\section{\nsRe~correlation} \label{correlation}

\begin{figure*}[ht!]
\includegraphics[width = \textwidth]{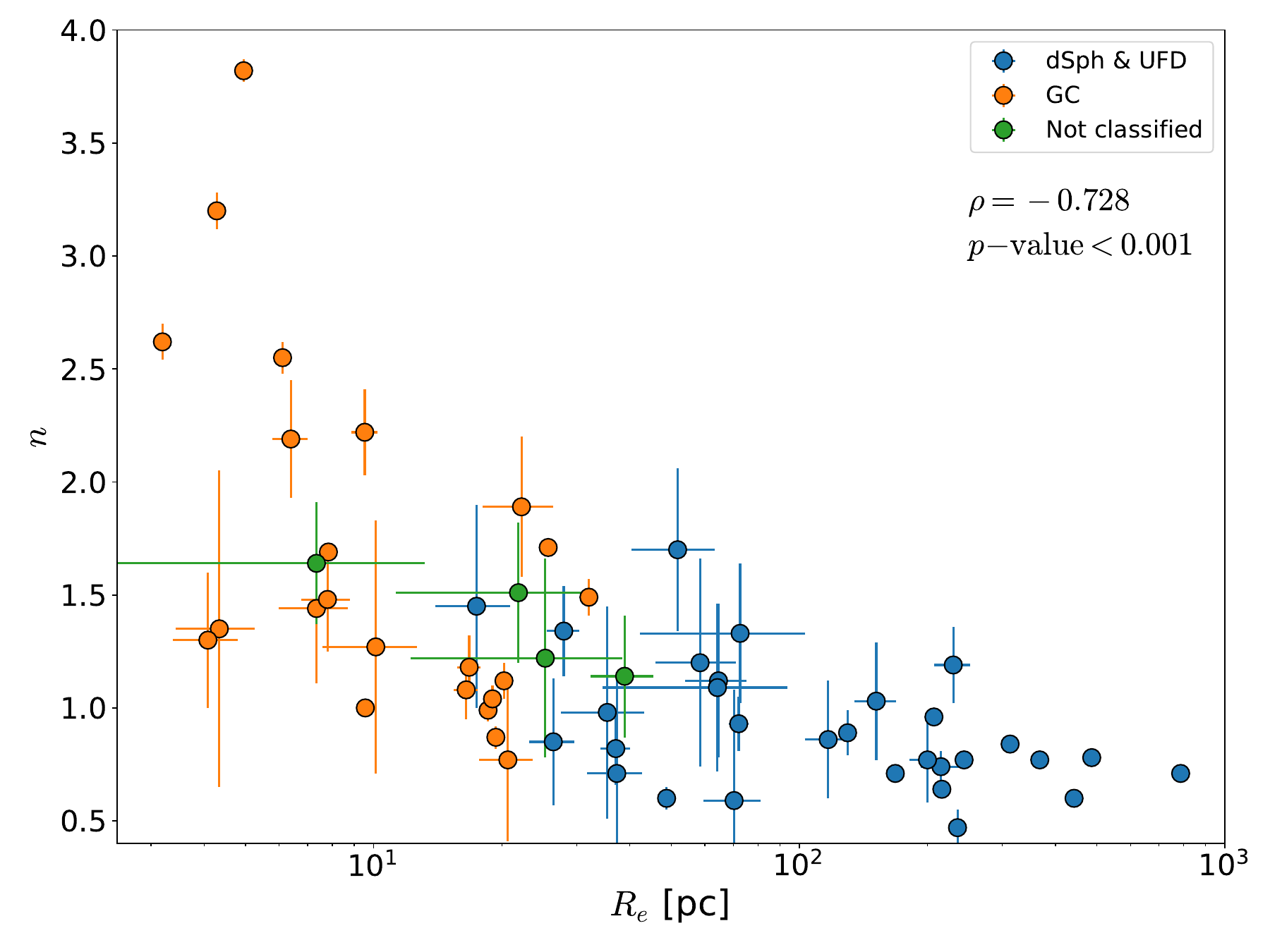}
\caption{Correlation between S\'ersic index and the effective radius in parsec for all the objects in our sample. \label{fig:sersic_effradius}}
\end{figure*}

An intriguing result regarding the overall properties of GCs and dwarf galaxies is a relation between the S\'ersic index and effective radii that is followed by all outer halo objects in our sample. In Figure \ref{fig:sersic_effradius} we present the relation between S\'ersic index and the effective radius for dwarf galaxies and GCs. The figure shows that the S\'ersic index decreases linearly with increasing size in log space.  This means that smaller satellite objects are more centrally concentrated than larger ones, since the S\'ersic index is a proxy for central concentration \citep{trujillo2001}. The Spearman's correlation coefficient is $-0.728$, with a p-value less than $0.001$, indicating that the correlation is significant at a high level.
A similar trend, but in the opposite sense has already been reported for dwarf and elliptical galaxies \citep{caon1993}, where larger galaxies have a higher S\'ersic index (i.e. are more concentrated). 

The fact that both GCs and dwarf galaxies share the same locus, forming a continuous group,
suggest a remarkable similarity between this two types of objects that is at first glance surprising, given that GCs and dwarf galaxies do not follow continuous trends in other structural parameter planes.

We note that, even though our structural parameters come from fitting a S\'ersic profile to the number density profiles, \citet{trujillo2001} demonstrated that a relation between \effrad~and \ns~cannot be produced by parameter coupling due to model fitting. 

\section{Discussion} \label{discussion}

\subsection{Origin of the \nsRe~relation}

The observed \nsRe~trend for GCs and dwarf galaxies does not have an obvious interpretation, especially if one takes into account that these objects have long been considered to be intrinsically different; GCs are believed to be dark matter-free while dwarf galaxies have been found to be heavily dark matter-dominated.

\citet{Graham2011} showed that it is possible to understand the existence of a relationship between the effective radius and mean effective surface brightness for elliptical and dwarf elliptical galaxies by showing that it can naturally arise if \muMv~and $\log(n) - M_V$ behave linearly, when both types of galaxies follow a S\'ersic density profile. In what follows, we 
consider a similar approach to understand the \nsRe~trend and
show an analytic procedure to reproduce the \nsRe~relation by considering linear fittings to the \muMv~and \nsMv~plots. 

Note that in \citet{Graham2011} they aim to explain a correlation in a different parameter space than \nsRe~(the one we present in this paper). However, the analytical procedure is the same. Unlike the case of elliptical galaxies, we know beforehand that GCs and dwarf galaxies do not form a single relation in either \muMv~and \nsMv~spaces and thus we follow \citet{Graham2011}'s procedure to investigate how the different behaviors in these parameter spaces can still result in the \nsRe~trend we detected. Additionally, we use the form \nsMv~instead of a $\log(n) - M_V$ relation. We do this because our range of \ns~is small enough that transforming to log space would not produce any substantial improvement. Moreover, by using the \nsMv~relation we avoid introducing an extra $\log(n)$ term in equation \ref{eq:theoretical_relation}.

The intensity profile at any given radius $r$ is modeled by the S\'ersic profile as:

\begin{equation}
I(r) = I_e \exp \left\{ -b_n \left[ \left( \frac{r}{r_e}\right)^{1/n} - 1\right]\right\}
\end{equation}

\noindent where $I_e$ is the intensity at the effective radius $r_e$, $n$ is the S\'ersic index and $b_n$ is a function that depends on $n$. As demonstrated by \citet{Graham2005}, from a S\'ersic profile it is possible to derive the following expression:

\begin{equation} \label{eq:Mtot}
M_\mathrm{tot} = \mu_\mathrm{e} - 2.5\log[f(n)] - 2.5\log(2\pi R_\mathrm{e,kpc}^2) - 36.57
\end{equation}

\noindent where $M_\mathrm{tot}$ is total absolute magnitude, \effmu~is the effective surface brightness, $R_\mathrm{e,kpc}$ is the effective radius in kiloparsecs, $n$ is the S\'ersic index and

\begin{equation*}
f(n) = \frac{n\mathrm{e}^b}{b^{2n}}\Gamma(2n)
\end{equation*}

\noindent with $b = 1.9992n - 0.3271$ for  $0.5 < n < 10$ \citep{Capaccioli1989} and $\Gamma$ is the gamma function.

Finally, if we consider relationships of the form

\begin{equation}
\label{eq:centmu-Mv}
\mu_\mathrm{0} = \mathrm{A}M_\mathrm{tot} + \mathrm{B}
\end{equation}

\begin{equation}
\label{eq:ns-Mv}
\ns = \mathrm{C}M_\mathrm{tot} + \mathrm{D}
\end{equation}

\noindent where \centmu~is the central surface brightness, and the fact that $\effmu = \centmu + 1.086b$, one obtains an equation that relates the S\'ersic index with the effective radius, of the form

\begin{equation}
\label{eq:theoretical_relation}
\log (R_\mathrm{e,kpc}) = \mathrm{E}\ns +  \mathrm{F}\log\left[f(\ns)\right] + \mathrm{G}
\end{equation}

\noindent with $\mathrm{E} = \frac{\mathrm{A}-1}{5\mathrm{C}} + 0.434$, $\mathrm{F} = -0.5$ and $\mathrm{G} = \frac{\mathrm{B}}{5} - \frac{\mathrm{D}(\mathrm{A}-1)}{5\mathrm{C}} - 7.784$.

This procedure shows that, for a S\'ersic density profile, 
linear relations in the \muMv~and \nsMv~spaces reproduce a relation in the \nsRe~space. One can reproduce other relationships between pairs of structural parameters if other linear relationships exist.

\subsection{Surface brightness vs. absolute magnitude}

\begin{figure*}
\includegraphics[width = 0.5\textwidth]{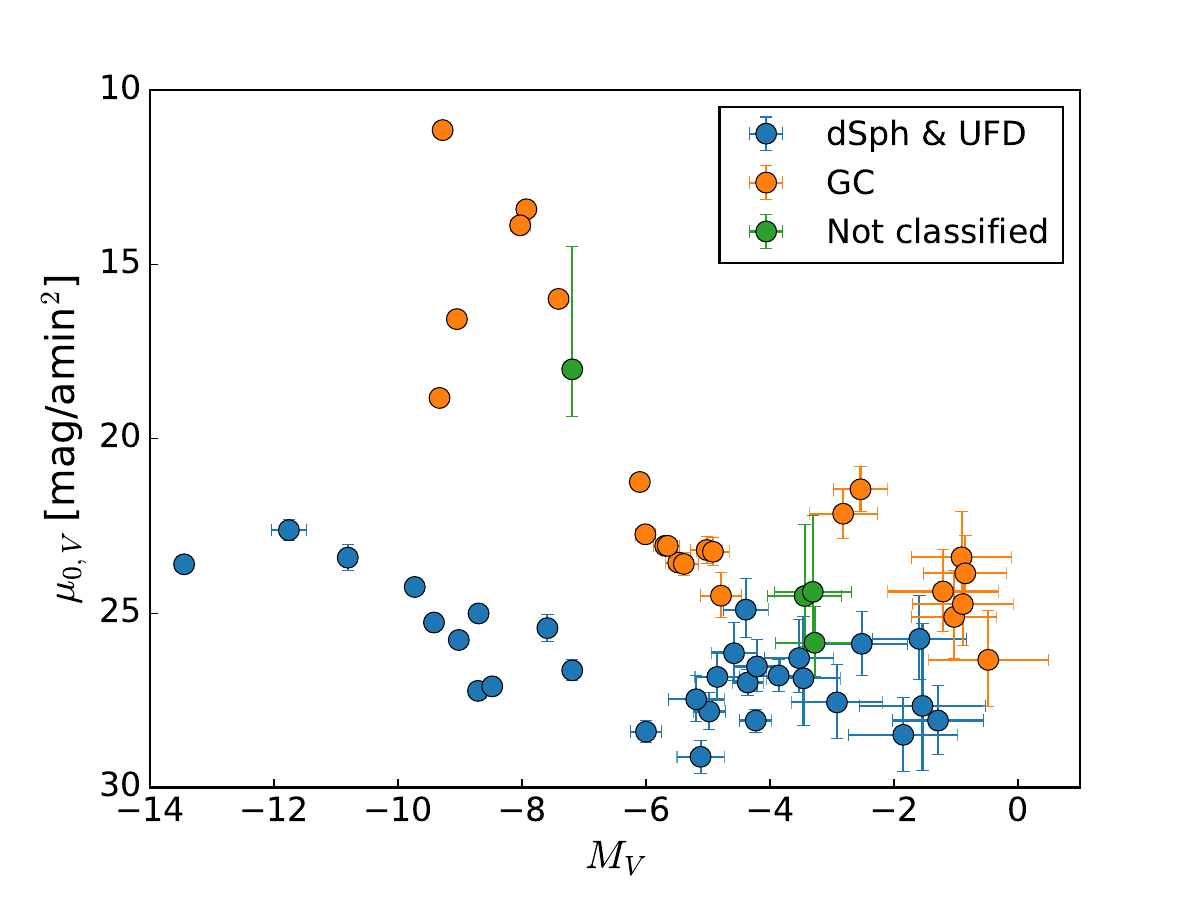}
\includegraphics[width = 0.5\textwidth]{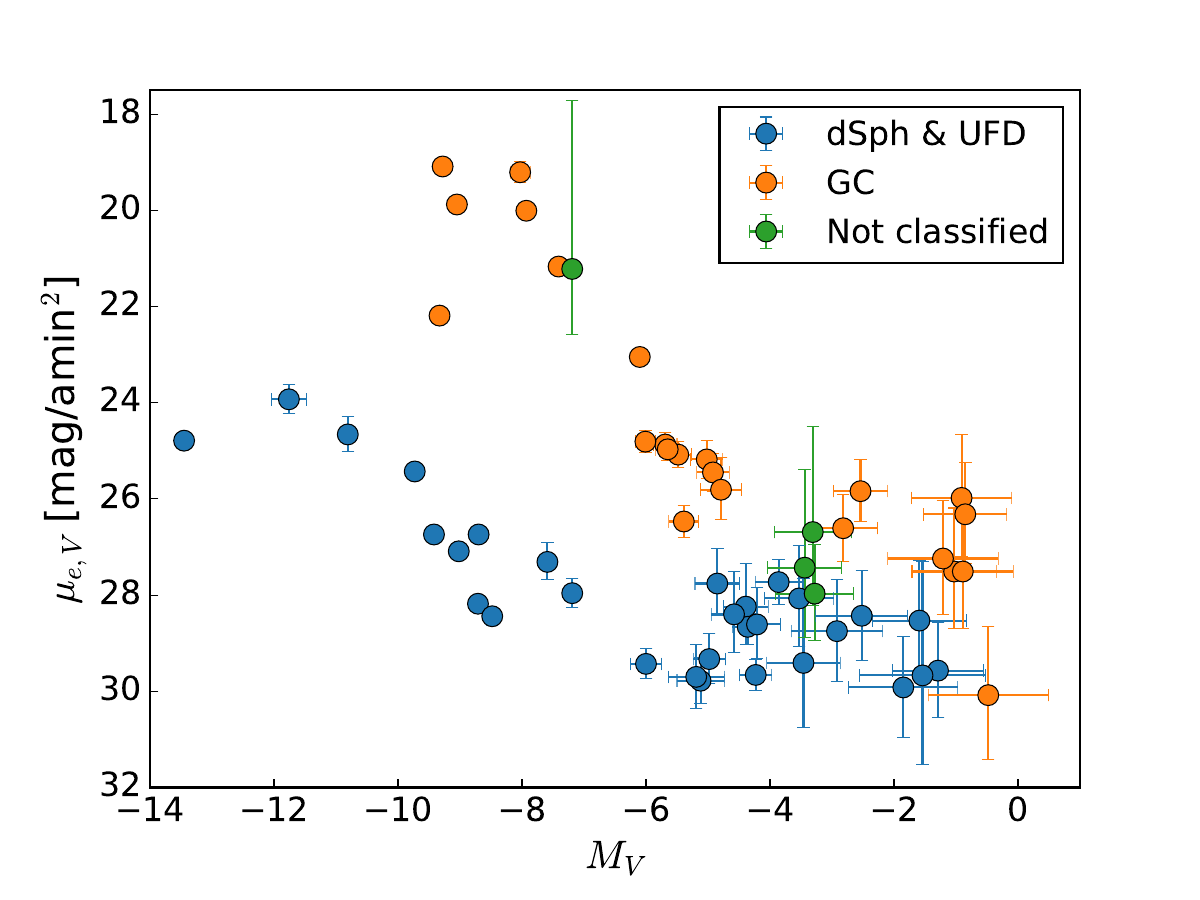}
\caption{Comparison of surface brightness against absolute magnitude for the objects in our dataset. \textit{Left panel:} Central surface brightness vs. absolute magnitude. \textit{Right panel:} Effective surface brightness vs. absolute magnitude. Note that in both panels it is evident that dwarf galaxies and globular clusters are well separated at high luminosities ($\Mv < \sim 5$). At lower luminosities both groups tend to mix, although on average globular clusters still show higher surface brightnesses. \label{fig:sbmag}}
\end{figure*}

Figure \ref{fig:sbmag} shows central and effective surface brightness versus absolute magnitude. In both plots, galaxies form a continuous group characterized by a luminosity v/s surface brightness dependency that flattens at $\Mv \sim -6$, in the region dominated by UFDs. This flattening was already identified by \cite{mcconnachie2012}, and it is possibly due to a detection bias, since the surface brightness of the least luminous UFDs are very near the detection limit of current surveys.

In the case of GCs, they show a higher central and effective surface brightness than galaxies at high luminosities (lower than $\Mv \sim -4$). At lower luminosities, GCs tend to concentrate at an almost constant surface brightness value, showing a similar behavior than the UFDs.

Although GCs and dwarf galaxies come closer at low luminosities in the \muMv~space, both groups do not mix completely; GCs have a higher average surface brightness than UFDs. This is not easily explained as a detection bias, since the surface brightness values at which GCs concentrate are higher than the detection limits.
A possible explanation for this different surface brightness floor is the fact that UFDs are believed to be currently embedded in a DM halo. 
For a given luminosity (or stellar mass), an object inside a DM halo
is likely more robust to tidal disintegration than a dark matter-free one and thus could reach lower luminosities, allowing also for lower surface brightnesses.

\subsection{S\'ersic index vs Absolute magnitude}

\begin{figure}[ht!]
\includegraphics[width = 0.5\textwidth]{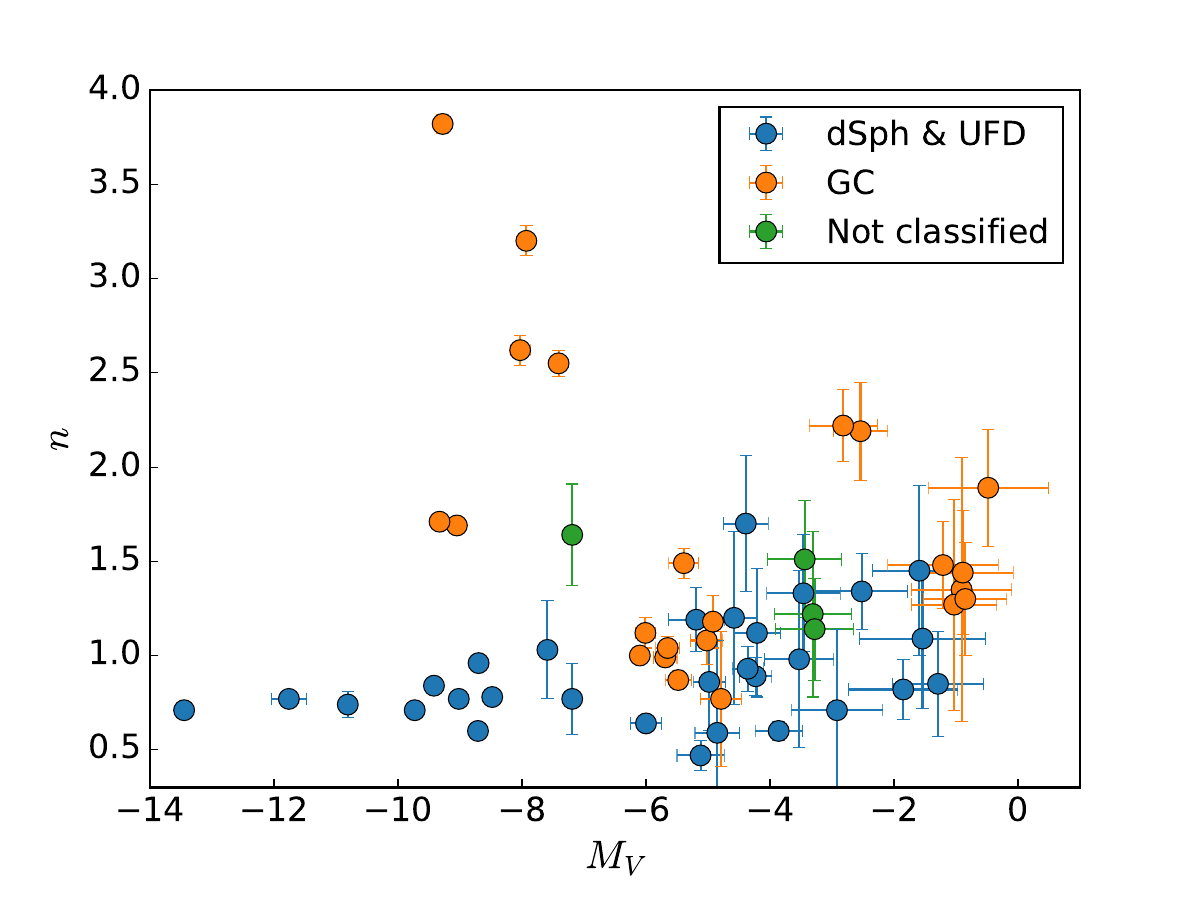}
\caption{S\'ersic index against absolute magnitude for all the objects in our dataset. Dwarf galaxies and globular clusters follow a nearly linear relationship that spans the whole range of luminosity, where low luminosity objects have a slightly higher S\'ersic index. There are six globular clusters located between $\Mv \sim -10$ and $\Mv \sim -8$ that do not follow this tendency, having high S\'ersic indexes for their luminosities. \label{fig:sersic_magnitude}}
\end{figure}

Figure \ref{fig:sersic_magnitude} shows the relation between the S\'ersic index and the absolute magnitude. Dwarf galaxies concentrate at relatively small values of the S\'ersic index, between $\sim 0.5$ and $\sim 1.5$, following a trend where the S\'ersic index increases slightly at lower luminosities. On the other hand, GCs do not seem to follow a single trend. Overall it appears that the S\'ersic index increases with luminosity. However, the data allow a different interpretation: most low luminosity GCs follow the trend delineated by dwarf galaxies, and only the six brighter clusters are off this trend and occupy a different region in the plot. 
In the surface brightness v/s  absolute magnitude relations (Figure \ref{fig:sbmag}), these clusters are also the ones with the highest surface brightnesses. In fact, from this Figure \ref{fig:sbmag}, it is also possible to infer that the low luminosity outer halo GCs and dwarf galaxies constitute a single group (although GCs have a higher mean central and effective surface brightnesses) with the high surface brightness GCs being outliers which may be part of a different subgroup of clusters.

In the next subsections, we apply to our dataset the procedure from which the \nsRe~relation originates, in order to see if it can be reproduced by linear fits obtained from the \muMv~and \nsMv~plots.

\subsection{Linear fits to \muMv~and \nsMv~relations}

\begin{figure*}[ht!]
\includegraphics[width = \textwidth]{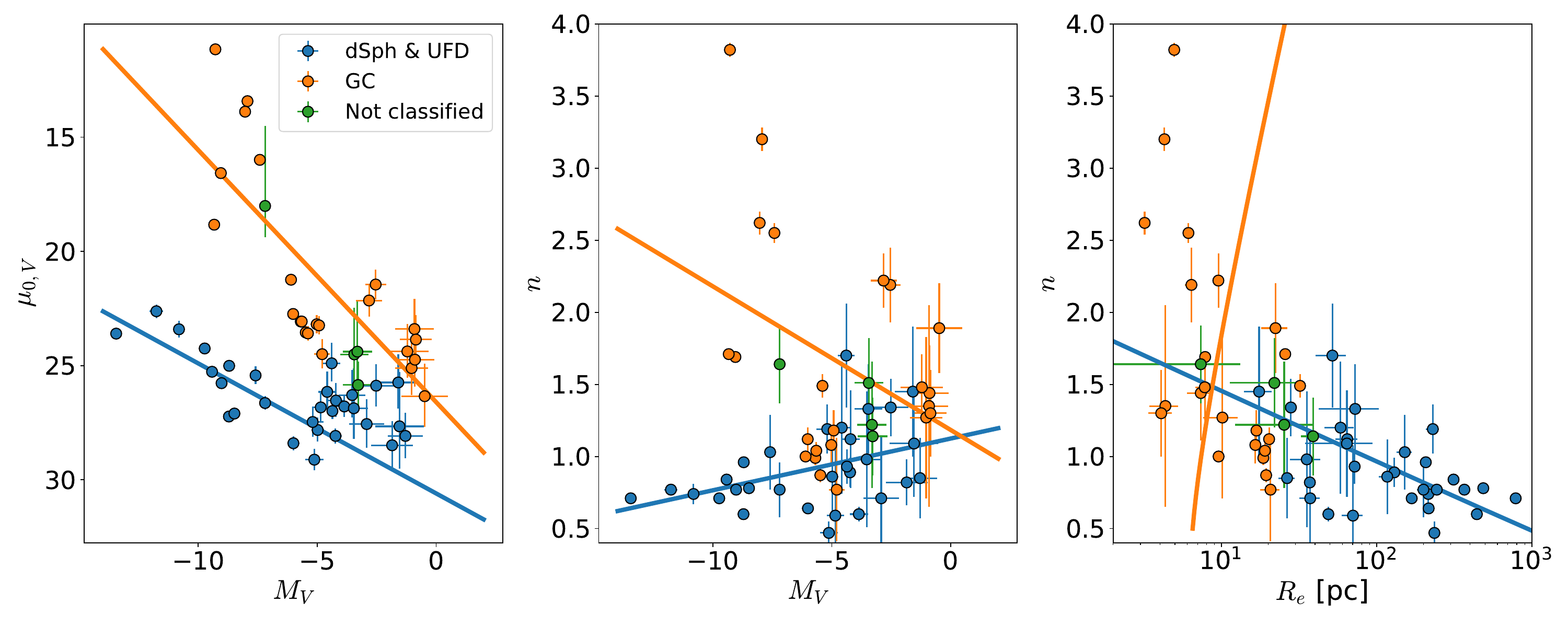}
\caption{Linear relation fits for the \muMv, \nsMv~plots and predicted relations for the \nsRe~correlation, for outer halo GCs and dSphs. \textit{Left} and \textit{Middle}: Linear relations fits for the \muMv~and \nsMv~plots, respectively. GCs are represented by orange circles, dSphs and UFDs are represented by blue circles and green circles represent not classified objects. The solid blue (orange) line represents the linear fit for dSphs (GCs). \textit{Right}: Solid lines represent the predicted relations for the \nsRe~correlation for GCs and for dwarf galaxies. Colors and symbols follow the same convention as the previous panels. \label{fig:linear_fits}}
\end{figure*}

As a first approach, we fit linear relations to dwarf galaxies and GCs assuming that they constitute two separate groups, following the conventional classification for each object. The left panel of Figure \ref{fig:linear_fits} shows the linear fit to the central surface brightness v/s absolute magnitude relation for GCs and galaxies. The fit for galaxies is given by

\begin{equation}
\label{eq:linear_fit_dsphs_sbmag}
\centVmu =  (0.569 \pm 0.120) \Mv + (30.597 \pm 1.105)
\end{equation}

\noindent while for GCs the fit is given by

\begin{equation}
\label{eq:linear_fit_gcs_sbmag}
\centVmu =  (1.104 \pm 0.194) \Mv + (26.598 \pm 1.091)
\end{equation}

Next, we analyze the relationship between the S\'ersic index and absolute magnitude for dwarf galaxies, which is presented in the middle panel of Figure \ref{fig:linear_fits}. The relations are characterized by

\begin{equation}
\label{eq:linear_fit_dsphs_sersicmag}
\ns = (0.036 \pm 0.015) \Mv + (1.124 \pm 0.101)
\end{equation}

\noindent while for GCs the fit is given by
\begin{equation}
\label{eq:linear_fit_gcs_sersicmag}
\ns = (-0.100 \pm 0.053) \Mv + (1.184 \pm 0.300)
\end{equation}

Finally, we obtain two relations similar to equation \ref{eq:theoretical_relation}, one for dwarf galaxies and another for GCs. For galaxies, the relation is

\begin{equation}
\begin{split}
\label{eq:theoretical_relation_dsphs}
\log_{10} (R_\mathrm{e,kpc}) & = (-1.965 \pm 1.211) \ns + -0.500\log_{10}(f(\ns)) \\& + (1.033 \pm 1.401)
\end{split}
\end{equation}

\noindent while for GCs the relation is
\\
\begin{equation}
\begin{split}
\label{eq:theoretical_relation_gcs}
\log_{10} (R_\mathrm{e,kpc}) & = (0.226 \pm 0.406) \ns + -0.500\log_{10}(f(\ns)) \\& + (-2.218 \pm 0.531)
\end{split}
\end{equation}

These derived relations are overplotted to our data in the right panel of Figure \ref{fig:linear_fits}. Dwarf galaxies seem to follow the predicted relation, represented by a blue solid line. On the contrary, GCs do not follow their predicted relation, represented by an orange solid line. This shows that separating our data in two groups, one composed by GCs and the other by dSphs, and fitting linear relations in the \muMv~and \nsMv~parameter spaces do not explain completely the observed correlation observed in the \nsRe~parameter space. This is expected for GCs, since it is clear that a linear fit in the \nsMv~for these objects is not a good model.

\subsection{Two separate globular cluster populations}

It is interesting that some of the GCs seem to follow the extrapolation of the  \nsRe~relation for dwarf galaxies. This prompts us to revisit the idea of 
two different GCs groups, and consider the possibility that some outer halo GCs do not constitute a different group from UFDs.

To further explore the origin of the \nsRe~correlation, we add to our sample the inner GCs data from \cite{Carballo2012}, covering a range in galactocentric radius from 11 to 21~kpc. We estimated their S\'ersic index and effective radius by fitting a S\'ersic profile to radial density profiles through a MCMC fitting procedure, where the free parameters are the S\'ersic index, the effective radius and the central surface density and used flat priors to estimate them. Given a degeneracy when estimating the S\'ersic index and the background surface density, we fixed the latter by visually exploring the density profiles for each inner cluster. We also obtained central surface brightness and absolute magnitude values from \cite{Harris1996} (2010 edition). Additionally, we add parameters of the object Kim 1 from the DES dataset, which where calculated in \citet{munoz17b}. It is relevant to mention that adding the inner halo GCs to the datasets breaks its homogeneity. However, this only affects the HSB group, keeping the homogeneity for the LSB clusters + galaxies group intact. 

\begin{figure*}[ht!]
\centering
\includegraphics[width = \textwidth]{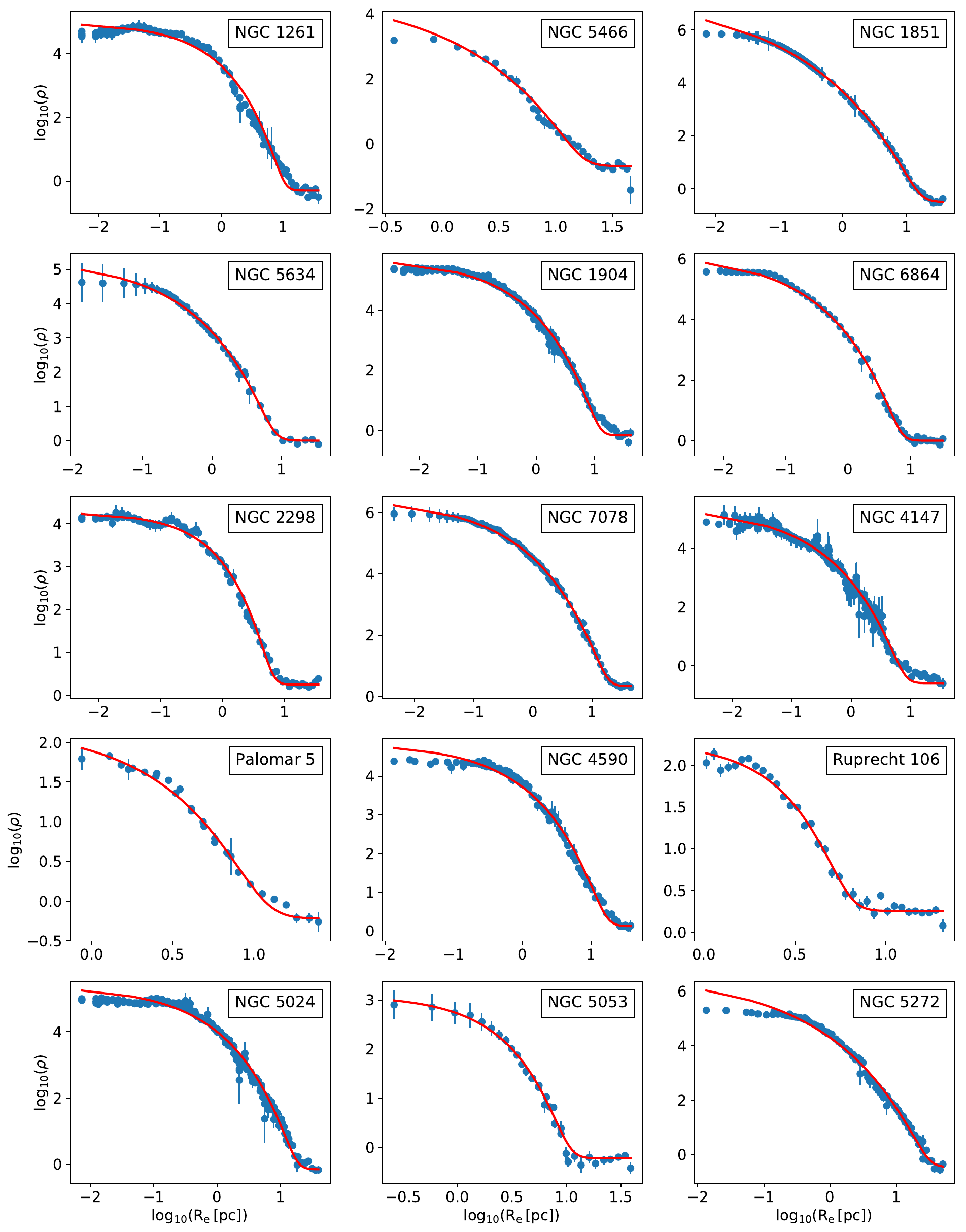}
\caption{Radial density profiles (blue circles) and fitted S\'ersic profile (red line) for each of the inner halo GCs from \citet{Carballo2012}. \label{fig:inner_fits}}
\end{figure*}

\begin{figure*}[ht!]
\centering
\includegraphics[width = \textwidth]{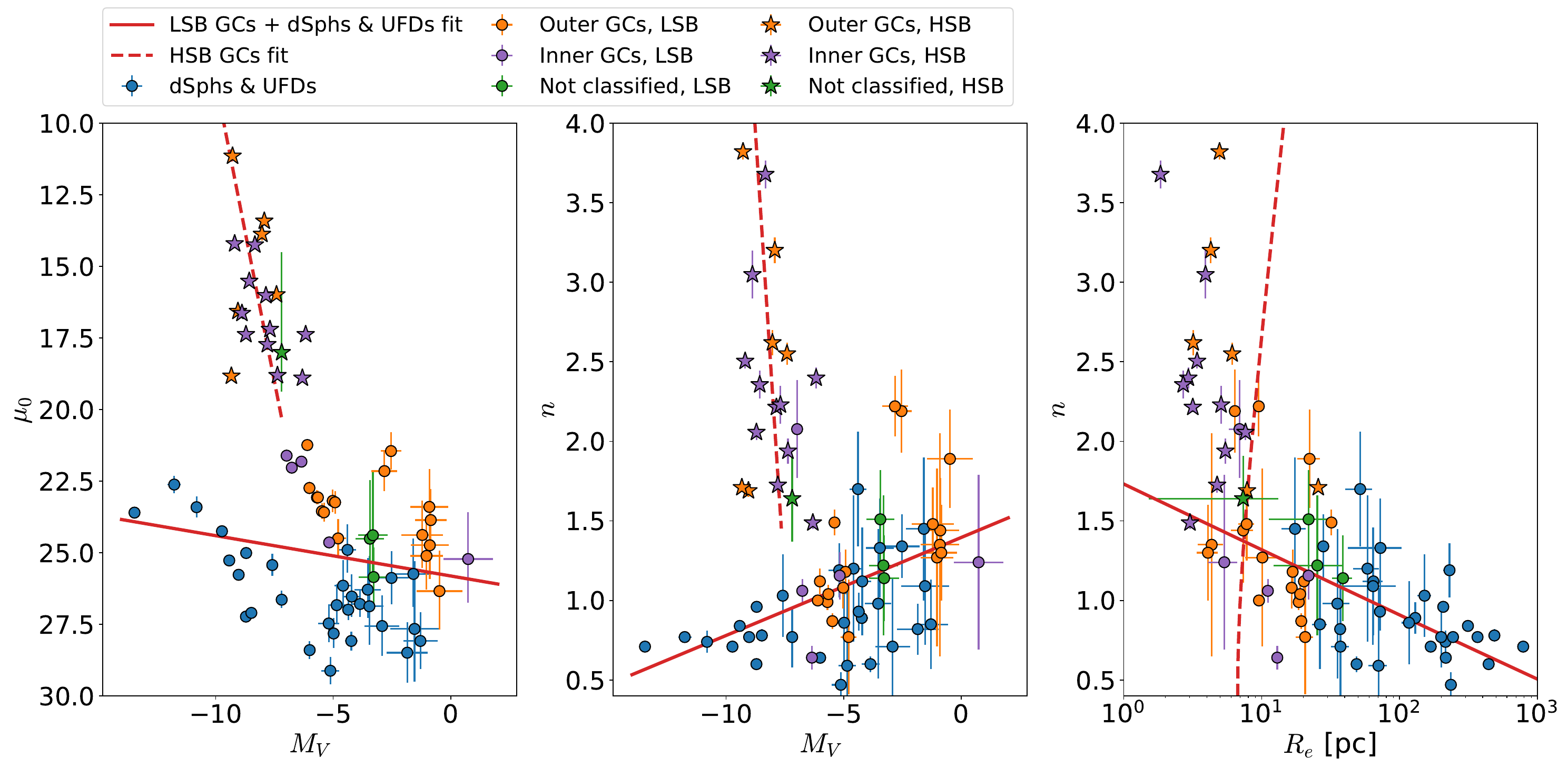}
\caption{Linear relation fits for the \muMv, \nsMv~plots and predicted relations for the \nsRe~correlation, for outer and inner halo GCs and dSphs. Color coding is the same as Figure \ref{fig:linear_fits}. Purple represents inner halo GCs. Stars represent satellite objects with $\centVmu < 20$ (HSB clusters), while circles are objects with $\centVmu \geq 20$ (LSB clusters plus dwarf galaxies). In the left and middle panels, the red, solid line represents the empirical linear fit to the LSB clusters plus dwarf galaxies group, while the red, dashed line is the empirical linear fit to the HSB clusters. In the right panel, lines are predicted relations based on the empirical linear fits of the left and right panels. \label{fig:with_inner_gcs}}
\end{figure*}

Table \ref{tab:inner_gcs} shows the estimated parameters for inner GCs and Figure \ref{fig:inner_fits} shows our S\'ersic profile fit to the radial density profiles of inner halo GCs. Figure \ref{fig:with_inner_gcs} shows the same plots as Figure \ref{fig:linear_fits}, but this time including the inner GCs mentioned before. As can be seen in the central surface brightness v/s absolute magnitude plot (left panel), most inner GCs are located in a high luminosity, high central surface brightness area, in comparison to most outer halo clusters. There are four inner halo clusters (purple circles) separated from the main group of inner halo clusters, below the $\centVmu \sim 20$ line. Additionally, some outer halo clusters (orange stars) are mixed with the inner halo GCs and separated from the rest of the outer halo cluster population (orange circles), which is located close to the UFD group. The separation between the two GC groups seems to be also marked by the $\centVmu \sim 20$ line. We tentatively name the groups of clusters above this line the High Surface Brightness (HSB) group, while the group of clusters below this division line is the Low Surface Brightness (LSB) group.

\begin{deluxetable*}{lrrrr}
\tablecaption{Parameters for inner halo GCs \label{tab:inner_gcs}}
\tablewidth{0pt}
\tablehead{
\colhead{Object} & 
\colhead{\Mv} & 
\colhead{\centVmu} &
\colhead{\effrad} & 
\colhead{\ns}\\
\colhead{} & 
\colhead{} & 
\colhead{(mag$/\arcsec^2$)} & 
\colhead{(pc)} & 
\colhead{}\\
}
\decimals
\startdata
Kim 1 & 0.74 & 25.22 & $5.36 \pm 1.27$ & $1.24 \pm 0.55$ \\
NGC 1261 & -7.80 & 17.73 & $4.75 \pm 0.12$ & $1.73 \pm 0.05$ \\
NGC 1851 & -8.33 & 14.25 & $1.85 \pm 0.04$ & $3.68 \pm 0.09$ \\
NGC 1904 & -7.86 & 16.02 & $3.17 \pm 0.04$ & $2.21 \pm 0.03$ \\
NGC 2298 & -6.31 & 18.90 & $3.02 \pm 0.06$ & $1.49 \pm 0.04$ \\
NGC 4147 & -6.17 & 17.38 & $2.94 \pm 0.06$ & $2.40 \pm 0.07$ \\
NGC 4590 & -7.37 & 18.81 & $5.46 \pm 0.17$ & $1.94 \pm 0.08$ \\
NGC 5024 & -8.71 & 17.38 & $7.63 \pm 0.14$ & $2.06 \pm 0.05$ \\
NGC 5053 & -6.76 & 22.03 & $11.08 \pm 0.40$ & $1.06 \pm 0.08$ \\
NGC 5272 & -8.88 & 16.64 & $3.92 \pm 0.22$ & $3.05 \pm 0.15$ \\
NGC 5466 & -6.98 & 21.61 & $6.91 \pm 1.13$ & $2.08 \pm 0.31$ \\
NGC 5634 & -7.69 & 17.20 & $5.08 \pm 0.18$ & $2.23 \pm 0.12$ \\
NGC 6864 & -8.57 & 15.52 & $2.70 \pm 0.09$ & $2.36 \pm 0.09$ \\
NGC 7078 & -9.19 & 14.21 & $3.40 \pm 0.05$ & $2.50 \pm 0.05$ \\
Palomar 5 & -5.17 & 24.64 & $21.86 \pm 1.10$ & $1.16 \pm 0.15$ \\
Ruprecht 106 & -6.35 & 21.82 & $12.98 \pm 0.50$ & $0.64 \pm 0.07$ \\
\enddata
\end{deluxetable*}

The middle panel of Figure~\ref{fig:with_inner_gcs}, S\'ersic index v/s absolute magnitude, also shows the separation of the HSB and LSB groups. While the HSB  group is distributed between $-10 < \Mv < -5$ and $1 < \ns < 4$, the LSB group occupies the region $-4 < \Mv$ and $\ns \lesssim 2$ and mix with the low luminosity part of the dwarf galaxy group. 

Finally, in the \nsRe~plot, right panel  in Figure \ref{fig:with_inner_gcs}, the HSB clusters are concentrated in the upper left region of the plot and deviate from the main correlation composed by LSB clusters and dwarf galaxies.

Adding the inner halo clusters to our outer halo sample reinforces the notion that there might be two subgroups of satellite objects: one composed by high luminosity, high central surface brightness clusters from the MW's inner and outer halo, and another composed by dwarf galaxies and GCs of lower central surface brightness and in general lower luminosity, which share the same parameter space occupied by low luminosity galaxies. With this in mind, we explore if fitting different linear relations to both groups defined above will reproduce the distribution in the S\'ersic index v/s effective radius plot. 

To find the relation in the \nsRe~parameter space for LSB GCs + dwarf galaxies and HSB GCs groups, we repeat the same procedure as before. For the LSB GCs + dwarf galaxies, the equations of empirical linear fit are

\begin{equation}
\label{eq:linear_fit_LSB_dsph_sbmag}
\centVmu =  (0.141 \pm 0.091) \Mv + (25.811 \pm 0.524)
\end{equation}

\begin{equation}
\label{eq:linear_fit_LSB_dsph_srsicmag}
\ns = (0.062 \pm 0.016) \Mv + (1.396 \pm 0.092)
\end{equation}

\noindent and its \nsRe~relation is

\begin{equation}
\begin{split}
\label{eq:theoretical_relation_LSB_dsph}
\log_{10} (R_\mathrm{e,kpc}) & = (-2.356 \pm 0.783) \ns + -0.500\log_{10}(f(\ns)) \\& + (1.273 \pm 1.128)
\end{split}
\end{equation}

For the HSB GCs group, the empirical linear fit equations are

\begin{equation}
\label{eq:linear_fit_HSB_sbmag}
\centVmu =  (4.190 \pm 1.985) \Mv + (50.439 \pm 25.170)
\end{equation}

\begin{equation}
\label{eq:linear_fit_HSB_sersicmag}
\ns = (-2.261 \pm 1.675) \Mv + (-15.848 \pm 11.878)
\end{equation}

\noindent and its \nsRe~relation is

\begin{equation}
\begin{split}
\label{eq:theoretical_relation_HSB}
\log_{10} (R_\mathrm{e,kpc}) & = (0.152 \pm 0.273) \ns + -0.500\log_{10}(f(\ns)) \\& + (-2.168 \pm 7.435)
\end{split}
\end{equation}

To fit the \muMv~and \nsMv~relations for HSB clusters, 
we in practice consider the \centVmu~and \ns~parameters as independent variables and \Mv~as the dependent variable, since the distribution of clusters on those
parameter spaces is strongly vertical when assuming \Mv~as the independent variable. Then, we invert the equations to obtain the coefficients when \Mv~acts as the independent variable. 

The results of this last procedure are shown in Figure \ref{fig:with_inner_gcs}. Left and middle panels show the linear fits to the \muMv~and \nsMv~plots, respectively. In each parameter space, the red solid line is the best linear fit to the LSB GCs + dwarf galaxy group, while the red dashed line is the best linear fit to the HSB GCs group. The right panel shows the predicted relations for the two groups, according to equation \ref{eq:theoretical_relation}. Objects in the LSB GCs + dSphs group follow the predicted relation (red solid line). On the other hand, clusters in the HSB group do not follow the predicted relation for them (red dashed line). This is surprising, since the linear fits for the HSB group closely follow the objects' distributions. One possible explanation for this is that HSB clusters are not well described by a pure S\'ersic profile and so equation \ref{eq:Mtot} does not hold for the HSB clusters. Supporting this idea, Figure \ref{fig:inner_fits} shows that the S\'ersic fit to the radial density profile in most cases is not ideal, especially in the central part, where the S\'ersic profile is cuspy and the observed radial density follows a core profile. The same is true for the outer halo GCs in the HSB group (Palomar 2, NGC 2419, NGC 5694, NGC 5824, NGC 6229 and NGC 7006) (see from Figure 6 to 16 in \citet{munoz17b}).

The fact that the \nsRe~correlation can be reproduced by using the same empirical relations for the LSB GCs and the dSphs and that the HSB clusters move off this trend might hint to the existence of two different groups of MW's outer halo GCs, one composed by GCs with structural and photometric properties similar to dwarf galaxies, and one composed by some outer halo GCs with properties more similar to inner halo GCs.

\subsection{Dark Matter in Globular Clusters}

The continuity between GCs and dwarf galaxies in the \nsRe~plot, together with the overlap of LSB cluster with ultra-faint galaxies in the \nsMv~and \muMv~plots, shows that 
the photometric properties of an important number of MW's clusters, at least in this plane, seem indistinguishable from those of dwarf galaxies. This may point to a common formation process for this two type of objects. Since it is commonly accepted that dwarf galaxies are embedded in a DM halo in which they formed, it is perhaps tempting to assume that GCs are also contained and/or were formed inside DM minihalos.

This is not necessarily a controversial idea, since simulations of GCs forming and relaxing inside DM halos do exist and they reproduce properties observed in real GCs \citep{Mashchenko2005a}. Other simulations show that tidal effects of the host galaxy can remove a large amount of the original DM inside GCs \citep{Mashchenko2005b}. More recently \cite{Penarrubia2017} showed that stars ejected due to hard encounters in the central region of GCs embedded in a DM halo generate an envelope of gravitationally-bounded stars. Supporting this model, so-called extra-tidal stars have been observed in many GCs of the MW \citep[e.g.][]{Carballo2012,Carballo2018} and the Andromeda Galaxy \citep{Mackey2010}. Additionally, \cite{Ibata2013} concluded that the presence of DM cannot be ruled out from the outer parts of the cluster NGC 2419. 

\subsection{Possible Origin of the HSB and LSB globular cluster groups}

In a scenario where all GCs formed through the same process, naively one could expect a continuity in their photometric properties. The existence of HSB and a LSB groups challenges that notion.
Here, we postulate that a possible explanation for the existence of these two groups is the effect of different processes of secular evolution due to different environments, something that is possible if some GCs formed and evolved inside the MW's potential, while others formed inside external satellite galaxies with weaker potentials and were later stripped from them during the MW's hierarchical accretion stage.

\cite{Zinn1993} studied the Galactic clusters and found that they can be classified into three different groups, according to their metallicity and Horizontal Branch (HB) morphology. There is a metal-rich group ($\mathrm{[Fe/H]} > -0.8$) located in the bulge and disk of the Galaxy (the Bulge/Disk group, or BD group), while a more metal-poor group ($\mathrm{[Fe/H]} < -0.8$) is found in the Galactic halo. \citeauthor{Zinn1993} also found that the halo group  contains clusters that can have a redder or bluer HB morphology for the same metallicity. This is the known second parameter effect and can be attributed to the age of the clusters, with redder clusters being younger than bluer ones for the same metallicity.  This led to the definition of an Old Halo (OH) and a Young Halo (YH) groups, where the former formed in-situ during a dissipative collapse while the latter formed inside the potential of dwarf galaxies that were later accreated by the MW.

Later, \cite{Mackey2004} supported this view,
showing that the metallicities and HB morphologies of GCs confirmed to be members of the Large Magellanic Cloud, Small Magellanic Cloud, Fornax and Sagittarius galaxies are consistent with the YH group values (see their Figure 13).

With this in mind, we can explore if the classification of GCs into HSB and LSB is consistent with the existence of the Bulge-Disk (BD), Old Halo (OH) and Young Halo (YH) groups. We use Table~1 of \cite{Mackey2005} to obtain the classification in the Zinn scheme for the clusters in our sample. Nine of them, all part of the LSB group, are not listed in the table. Moreover, these clusters do not show any horizontal branch in the color-magnitude plot in \cite{munoz17b}, so is not possible to measure an HB index, necessary to classify them (Table \ref{tab:hb_class} shows this classification for our clusters present in \cite{Mackey2005}). Of the remaining clusters, we count 17 GCs in the HSB group  and 13 in the LSB group (in both cases including the inner halo GCs from \citet{Carballo2012}). According to the classification in \citet{Mackey2005}, of the HSB group 9 clusters are OH, 7 are YH and 1 is SG (part of the Sagittarius dwarf galaxy); in the LSB group, 2 are OH and 11 are YH. In other words, about half of the clusters in the HSB group are consistent with an in-situ origin and the other half are consistent with an external origin, while in the LSB group, the vast majority ($\sim 85\%$) are consistent with an external origin. 

\begin{deluxetable*}{lcc}
\tablecaption{HB classification for the cluster in our sample that have that information in \cite{Mackey2005}. YH stands for Young Halo cluster, OH for Old Halo cluster and SG for Sagittarius cluster.  \label{tab:hb_class}}
\tablewidth{0pt}
\tablehead{
\colhead{Object} & \colhead{Surface Brightness Class} & \colhead{HB Class}\\
}
\startdata
NGC 1261 & HSB & YH\\
NGC 1851 & HSB & OH\\
NGC 1904 & HSB & OH\\
NGC 2298 & HSB & OH\\
NGC 2419 & HSB & OH\\
NGC 4147 & HSB & SG\\
NGC 4590 & HSB & YH\\
NGC 5024 & HSB & OH\\
NGC 5272 & HSB & YH\\
NGC 5634 & HSB & OH\\
NGC 5694 & HSB & OH\\
NGC 5824 & HSB & OH\\
NGC 6229 & HSB & YH\\
NGC 6864 & HSB & OH\\
NGC 7006 & HSB & YH\\
NGC 7078 & HSB & YH\\
Palomar 2 & HSB & YH\\
\hline
AM1 & LSB & YH\\
Eridanus & LSB & YH\\
NGC 5053 & LSB & YH\\
NGC 5466 & LSB & YH\\
NGC 7492 & LSB & OH\\
Palomar 13 & LSB & YH\\
Palomar 14 & LSB & YH\\
Palomar 15 & LSB & OH\\
Palomar 3 & LSB & YH\\
Palomar 4 & LSB & YH\\
Palomar 5 & LSB & YH\\
Pyxis & LSB & YH\\
\enddata
\end{deluxetable*}

To explain the current properties of HSB and LSB clusters, tidal stripping processes must have affected GCs differently. In light of the idea that OH clusters formed in-situ and YH clusters did so in external galaxies, it is evident that HSB and LSB clusters must have been affected by different tidal forces during their secular evolution because they were located at different galactic host environment. OH clusters, formed in-situ, were subjected to a stronger tidal force, stripping stars from the high luminosity clusters (this would originate OH clusters with HSB group characteristics) and completely disintegrated cluster of lower stellar mass (this would explain why there are almost no OH clusters with LSB group properties). YH clusters, on the other hand, formed in external galaxies with weaker potential, so they were affected by a weaker tidal force. Later, with the accretion of dwarf galaxies by the MW, they were incorporated to its GCs system. Some of these clusters have already been disrupted by the MW's tidal force, leaving a stream of stars behind \citep[e.g.][]{Grillmair2009}; others are in the process of disintegration, as evidenced by the tidal tails emerging from them \citep[e.g. Palomar~5,][]{Rockosi2002}; and others still survive because they have not been affected by the MW's tidal force long enough or do not live in destructive orbits. Among this last group, there are clusters of low luminosity and extended (the ones that constitute the LSB group) and others of higher luminosity and compact (characteristics of HSB GCs).

\citet{hurley2010}, through N-Body simulations, provided further insights of the formation of GCs in galactic gravitational potentials of different intensities. They showed that Large Magellanic Cloud-like galaxies of weak tidal fields can produce extended clusters of up to $30$\,pc from a standard process of formation and evolution. Furthermore, they show that, for GCs forming in MW-like tidal fields at 10\,kpc from the galactic center, their maximum half-light radius is $\sim 10$\,pc. Finally, they point that MW-like galaxies could form extended clusters at large galactocentric distances ($\sim100$\,kpc) and any extended cluster present at the inner portions of the galaxy likely formed inside an accreted dwarf galaxy. These simulations support the idea that LSB clusters (typically extended) formed in accreted dwarf galaxies, while HSB clusters (usually more compact) formed inside the MW.

The notion that LSB GCs are of external origin while HSBs are a mix of clusters formed in-situ and externally, could explain the differences presented in this work. To confirm or reject this idea, the best way is to know the orbit of each satellite object. However, this has proven to be a hard task, since to constrain their orbits it is necessary to perform high precision phase space measurements, something that is difficult in objects with a low number of member and/or low luminosity stars.

The \textit{Gaia} mission promises high precision kinematic information for many of the satellite objects. In fact, the second data release of this mission has already provided us with very accurate proper motions for some satellite dwarf galaxies \citep{GaiaCol2018, Fritz2018, Kallivayalil2018, Massari2018, Pace2018, Simon2018} and inner halo GCs \citep{GaiaCol2018, Vasiliev2018}. This new information has allowed to conclude that most of UFDs inside a galactocentric radius of 100\,kpc follow eccentric, high velocity and retrograde orbits and some of the galaxies are consistent with being in their first infall \citep{Simon2018}. In the case of GCs, \citet{Vasiliev2018} showed that clusters in $R_\mathrm{G} \lesssim 10\,\mathrm{kpc}$ rotate in prograde orbits and that the velocity dispersion is isotropic, while for clusters further out the velocity distribution becomes radially anisotropic.

In light of these results, we predict that the LSB GCs should follow orbits similar to UFDs' and that they are on their first infall. This last point is consistent with the existence of such low luminosity, low surface brightness objects inside the strong tidal field of the MW. Also, they should exhibit a radially anisotropic velocity distribution. For the HSB group, given its mixed composition, we predict that GCs located in the inner halo should follow prograde orbits (consistent with clusters formed in-situ) with an isotropic velocity dispersion, while the ones located near the frontier between inner and outer halo should have kinematics similar to the LSB clusters.

\section{Conclusions} \label{conclusions}

In this work, we explored in detail a strong correlation between the S\'ersic index and half-light radius that is followed by almost all the outer halo satellite objects included in our Megacan sample \citep{munoz17a}. More importantly, in this trend a large number of GCs follow the same locus as dwarf galaxies, adding support to the similarities between these two type of objects.

We followed the procedure of \citet{Graham2005} to see if the correlation in the \nsRe~plot can be a consequence of empirical linear relations in the \muMv~and \nsMv~parameter spaces for objects that follow a S\'ersic density profile. We showed that this is possible if we consider two different class of outer halo GCs: one that is composed by clusters of low surface brightness, with properties similar to ultra-faint dwarf galaxies (the LSB group); and another that is composed by clusters of high surface brightness, with properties similar to inner halo GCs (the HSB group). From our analysis, we saw that empirical linear relations can be fit to the LSB GCs + dwarf galaxies group and for the HSB GCs group (including the inner GCs). However, for HSB GCs the \nsRe~relation cannot be reproduced, probably because they are not fully described by a pure S\'ersic profile. 

Given the strong similarities between LSB GCs and dwarf galaxies, and considering that the latter are dominated by DM, we proposed that this is consistent with the notion that GCs also formed inside halos of DM, sharing a common formation process. This idea is supported by previous works that show that GCs with properties similar to what is empirically observed today can be originated thorough a formation process inside a DM halo.

Finally, to explain the existence of HSB and LSB GCs in a scenario were all clusters formed through a common process, we proposed that tidal effects of the host galaxy play a major role in shaping of cluster's properties. GCs of both types are formed inside MW-like and dwarf galaxies. However, the ones formed inside MW-like galaxies are subjected to stronger tidal forces than the ones inside dwarf galaxies. Thus, LSB GCs inside the MW were quickly disrupted, while HBS GCs, given their higher masses and densities, survived, albeit losing part of their mass; on the other hand, both HSB and LSB GCs survived inside dwarf galaxies. Later, during the process of accretion of dwarf galaxies by the MW, the external HSB and LSB GC populations were incorporated to our Galaxy's cluster system. From this moment, the stronger potential of the MW started its tidal effect over them. The scenario just proposed would explain the observed proportion of external and in-situ origin for both HSB and LSB GCs. In fact, following the classification scheme proposed by \citet{Zinn1993}, around half of the HSB GCs are of OH type, while the other half is of YH type, a distribution consistent with a mix of external and in-situ origin. In contrast, for the LSB group, almost all of them are of YH type, suggesting that the majority of them were stripped from accreted dwarf galaxies.

Future high precision proper motions measurements of satellite galaxies, especially for UFDs and outer halo GCs, will allow to know the true origin of HSB and LSB clusters. We predict that the majority of LSB clusters should have orbits similar to UFDs and dSphs, while HSB clusters should orbit the Galaxy in a way similar to inner halo GCs.

\vspace{0.5cm}
S.~M-L.~acknowledges support from CONICYT/Doc\-to\-ra\-do Na\-cio\-nal/2013-21130655. J.~A.~C-B.~ acknowledges financial support to CAS-CONICYT 17003. R.~R.~M.~acknowledges partial support from BASAL Project AFB-$170002$ as well as FONDECYT project N$^{\circ}1170364$.

\end{document}